\documentclass[prd,twocolumn,showpacs,nofootinbib]{revtex4}

\usepackage{times}
\usepackage{natbib}
\usepackage{epsfig}
\usepackage{amsmath}
\usepackage{bm}
\usepackage{comment}         
\usepackage{color}

\newcommand{\bdv}[1]{{\bf #1}}
\newcommand{\up}[1]{{\rm #1}}
\newcommand{\mpc}{\up{Mpc}}
\newcommand{\gpc}{\up{Gpc}}
\newcommand{\Xang}{\bdv{\hat x}}
\newcommand{\ket}[1]{|#1\rangle}
\newcommand{\xvec}{\bdv{x}}
\newcommand{\kvec}{\bdv{k}}
\newcommand{\beeq}{\begin{equation}}
\newcommand{\bra}[1]{\langle#1|}
\newcommand{\eneq}{\end{equation}}
\newcommand{\braket}[2]{\langle#1|#2\rangle}
\newcommand{\AVE}[1]{\langle#1\rangle}
\newcommand{\bear}{\begin{eqnarray}}
\newcommand{\enar}{\end{eqnarray}}
\newcommand{\Kang}{\bdv{\hat k}}

\newcommand{\SP}{\mathcal{S}}
\newcommand{\RW}{\mathcal{W}}
\newcommand{\PP}{\mathcal{P}_z}
\newcommand{\RR}{\mathcal{P}_r}
\newcommand{\TT}{\mathcal{T}}
\newcommand{\cp}{\varphi_v}         
\newcommand{\ST}{\mathcal{M}}
\newcommand{\NP}{\mathcal{N}}
\newcommand{\dv}{\delta_v}
\newcommand{\dz}{\delta z}           
\newcommand{\ddL}{\delta\mathcal{D}_L} 
\newcommand{\ax}{\alpha_{\chi}}     
\newcommand{\px}{\varphi_{\chi}}    
\newcommand{\dzg}{\dz_\chi}         
\newcommand{\HH}{\mathcal{H}}   
\newcommand{\drg}{\delta\mathcal{R}}  
\newcommand{\kag}{\mathcal{K}}     
\newcommand{\OM}{\Omega_m}
\newcommand{\dm}{\delta_m}
\newcommand{\vx}{v_{\chi}}         
\newcommand{\WW}{W}
\newcommand{\TTm}{\mathcal{T}_m}
\newcommand{\PZ}{P_z}
\newcommand{\hmpc}{{h^{-1}\mpc}}
\newcommand{\PM}{P_m}
\newcommand{\RA}{\rightarrow}

\newcommand{\lmax}{l_\up{max}}
\newcommand{\kmin}{k_\up{min}}
\newcommand{\kmax}{k_\up{max}}
\newcommand{\hgpc}{{h^{-1}\gpc}}
\newcommand{\FF}{\mathcal{F}}

\begin{document}

\title{All-sky analysis of the general relativistic galaxy power spectrum}

\author{Jaiyul Yoo$^{1,2}$}
\altaffiliation{jyoo@physik.uzh.ch}
\author{Vincent Desjacques$^3$}
\altaffiliation{Vincent.Desjacques@unige.ch}

\affiliation{$^1$Institute for Theoretical Physics, University of Z\"urich,
CH-8057 Z\"urich, Switzerland}
\affiliation{$^2$Lawrence Berkeley National Laboratory, University of 
California, Berkeley, CA 94720, U.S.A.}
\affiliation{$^3$D\'epartement de
Physique Th\'eorique and Center for Astroparticle Physics, Universit\'e de
Gen\`eve, CH-1211 Gen\`eve, Switzerland}

\date{\today}

\begin{abstract}
We perform an all-sky analysis of the general relativistic galaxy power 
spectrum using the well-developed spherical Fourier decomposition.
Spherical Fourier analysis expresses the observed galaxy fluctuation
in terms of the spherical harmonics and spherical Bessel functions that are
angular and radial eigenfunctions of the Helmholtz equation, providing
a natural orthogonal basis for all-sky analysis of the large-scale mode
measurements. Accounting for all the
relativistic effects in galaxy clustering, we compute the spherical power 
spectrum and its covariance matrix and compare it to the standard
three-dimensional power spectrum to establish a connection.
The spherical power spectrum recovers the three-dimensional 
power spectrum at each wavenumber~$k$
with its angular dependence $\mu_k$ encoded in angular multipole~$l$,
and the contributions of the line-of-sight projection to galaxy clustering
such as the gravitational lensing effect can be readily accommodated
in the spherical Fourier analysis.
A complete list of formulas for computing the relativistic spherical galaxy 
power spectrum is also presented.
\end{abstract}

\pacs{98.80.-k,98.65.-r,98.80.Jk,98.62.Py}

\maketitle

\section{Introduction}
\label{sec:intro}

The past few decades have seen a rapid progress in large-scale galaxy surveys.
The Sloan Digital Sky Survey (SDSS; \cite{YOADET00}) and the Two degree
Field Galaxy Redshift Survey (2dFGRS; \cite{CODAET01}) opened a new horizon
in modern cosmology, mapping three-dimensional positions of millions of 
galaxies. The Baryonic Oscillation Spectroscopic Survey (BOSS; \cite{SCBLET07})
and the Dark Energy Survey (DES; \cite{DESC05}) represent the current
state-of-the-art galaxy surveys, together with the recently completed
WiggleZ Dark Energy Survey \cite{DRJUET10}.
The exploration of the large scale structure of the Universe will 
continue with
future galaxy surveys such as Euclid,\footnote{\scriptsize http://sci.esa.int/euclid}
the BigBOSS,\footnote{\scriptsize http://bigboss.lbl.gov}
the Large Synoptic Survey Telescope,\footnote{\scriptsize http://www.lsst.org}
and the Wide-Field InfraRed Survey 
Telescope.\footnote{\scriptsize http://wfirst.gsfc.nasa.gov}
As they will cover a substantial fraction of the entire sky and a wide range
of redshift, these future surveys will measure galaxy clustering with stupendous
statistical power, demanding thereby that the current theoretical predictions
be refined to achieve higher levels of accuracy.

The galaxy distribution measured in galaxy surveys represents a biased version 
of the underlying matter distribution.
A traditional approach to analyzing galaxy number density fields
is to utilize the power spectrum of its Fourier components.
Since each Fourier mode evolves independently in the linear regime,
the power spectrum measurements can be used to infer the primordial matter
power spectrum and to extract cosmological information (see
\cite{TEBLET04,PENIET07,REPEET10} for recent power spectrum measurements).
While the power spectrum analysis merits its intuitively 
simple interpretation of the measurements in connection with the underlying
matter distribution, it assumes that the density fields are defined in a 
cubic volume. As the recent and forthcoming galaxy surveys cover a 
progressively larger fraction of the sky, the validity of the flat-sky 
approximation and the power spectrum analysis becomes questionable. 

At the same time, there exists a demand 
for large-scale 
measurements from the theoretical side. In the past few years 
the relativistic description of galaxy clustering has been
developed \cite{YOFIZA09,YOO10}. The advance in
theoretical development results from the finding that the observed
quantities in galaxy clustering such as the observed redshift and the galaxy 
position on the sky are different from quantities used to construct the
observed galaxy fluctuation such as the background redshift and the unlensed
galaxy position \cite{YOO09}.
Those theoretical quantities are gauge-dependent, and
the subtle difference in those quantities become substantial on large
scales, where the relativistic effect becomes important \cite{YOFIZA09,YOO10}. 
The full relativistic formula of galaxy clustering can be analytically
derived at the linear order in perturbations, providing a complete picture
of galaxy clustering on large scales 
\cite{YOFIZA09,YOO10,CHLE11,BODU11,BASEET11,BRCRET12,JESCHI12}. 
Furthermore, it is shown \cite{YOHAET12} that these relativistic effects in
galaxy clustering can be measured in future galaxy surveys, providing
a great opportunity to test general relativity on cosmological scales.

The relativistic formula for the observed galaxy fluctuation 
$\delta^\up{obs}_g(z,\Xang)$ is well-defined in observations, 
where it is a function of the observed redshift~$z$ and 
angle~$\Xang$ on the sky. Since 
the wide angle formalism \cite{SZMALA98,SZAPU04,MATSU00b,PASZ08}
has been developed to compute the two-point correlation function of 
the full Kaiser formula without adopting the distant-observer approximation,
its extension to the full relativistic formula can be readily 
made \cite{MODU12,BEMAET12}. 
However, the resulting equation for the two-point correlation function
is highly complicated, even for the full Kaiser formula, mainly due to
the geometrical effect. Furthermore, its physical interpretation is
not as straightforward as in the power spectrum analysis owing to its 
nontrivial relation to the primordial matter power spectrum. 

Adopting the flat-sky approximation, \citet*{YOO10} computed the galaxy power 
spectrum, accounting for the relativistic effects (see, also, \cite{JESCHI12}).
While it is always possible to embed the observed sphere in a cubic volume with
rectangular coordinates and to perform a power spectrum analysis, it becomes
difficult in principle to connect these large-scale 
measurements to the underlying theory, because 
the flat-sky
approximation has a limited range of validity.
However, it is shown \cite{YOSE13} that, on these large scales where 
measurement uncertainties are significant, 
the systematic errors associated with 
the flat-sky approximation are indeed negligible in the power spectrum 
analysis, {\it if performed properly}.
This is consistent with the previous finding of \cite{YOO10,YOHAET12} obtained 
by a simpler treatment.

Here we present an alternative to the traditional power spectrum analysis,
based on radial and angular eigenfunctions of the Helmholtz equation.
The spherical Fourier analysis has been well developed 
\cite{BIQU91,FILAET95,HETA95} in galaxy clustering, 
while its application was limited to the Kaiser formula. The observed galaxy 
fluctuation is decomposed in terms of Fourier modes 
and spherical harmonics, both of which 
provide a natural orthogonal basis for all-sky analysis.
In observational side, \citet{FISCLA94} applied the spherical Fourier
analysis to the IRAS Redshift Survey, and the method was further
developed in \citet{FILAET95} to reconstruct the velocity and the gravitational
potential fields. Theoretical refinement has been made in the past few years.
\citet{RARE12} computed the spherical power spectrum of the matter density
with focus on the baryonic acoustic oscillation feature, ignoring the
redshift-space distortion (see \cite{PRMU13} for the effect of the
redshift-space distortion and nonlinearity).
\citet{SHCRPE12} used the spherical Fourier
analysis of the redshift-space distortion and its cross-correlation with
the CMB temperature anisotropies to constrain the growth rate of structure.
The same technique is also applied to the weak gravitational lensing formalism
in \citet{HEAVE03} to take advantage of the information on distances to
background source galaxies. Compared to the traditional weak lensing,
this spherical Fourier analysis is known as the 3D weak lensing, and it is
shown in \citet{KIHEMI11} that the 2D tomography in weak lensing is just the 3D
weak lensing with the Limber approximation. 

Drawing on this previous work,
we perform a spherical Fourier analysis of the observed galaxy clustering.
Accounting for all the general relativistic effects in galaxy
clustering, we compute the spherical power spectrum and its covariance matrix
and compare them to the corresponding three-dimensional power spectrum.
The organization of this paper is as follows. In Sec.~\ref{sec:for}, we 
briefly review the spherical Fourier analysis. We first discuss the basic
formalism of the spherical Fourier analysis in Sec.~\ref{ssec:decomp}
and its application to galaxy redshift survey in Sec.~\ref{ssec:survey}.
In Sec.~\ref{sec:gcgr}, we present the full general relativistic description
of galaxy clustering. We first describe the key equations behind the 
relativistic description in Sec.~\ref{ssec:gr} and present its relation 
to the spherical Fourier analysis in Sec.~\ref{ssec:weight}. In 
Sec.~\ref{ssec:limber}, we provide the Limber approximation to the
spherical Fourier analysis of the relativistic description. 
The main results of the spherical Fourier analysis are presented in 
Sec.~\ref{sec:results}, in which we discuss the spherical power spectrum of the
relativistic formula in Sec.~\ref{ssec:spow} and the measurement uncertainties
associated with them in Sec.~\ref{ssec:uncer}. Finally, we discuss the
implications of our results in Sec.~\ref{sec:discussion}.
Our numerical calculations are performed by assuming a flat $\Lambda$CDM 
universe with cosmological parameters consistent with the WMAP7 
results \cite{KOSMET11}.

\section{Formalism}
\label{sec:for}
Spherical Fourier analysis of three-dimensional density fields has been
well developed \cite{BIQU91,FILAET95,HETA95} (see, also, \cite{HACU96} for a
different approach). 
We briefly review the basics of spherical Fourier decomposition 
in Sec.~\ref{ssec:decomp} and discuss its application to galaxy surveys
in Sec.~\ref{ssec:survey}, accounting for issues associated with redshift
distribution.

\subsection{Spherical Fourier Decomposition}
\label{ssec:decomp}
We begin by defining our normalization convention for Fourier decomposition.
The position vector $\ket{\xvec}$ and the Fourier
mode $\ket{\kvec}$ are normalized to satisfy the completeness relation:
\beeq
1=\int d^3\xvec~\ket{\xvec}\bra{\xvec}
=\int {d^3\kvec\over(2\pi)^3}~\ket{\kvec}\bra{\kvec}~.
\eneq
With the plane wave represented by
\beeq
\braket{\xvec}{\kvec}\equiv\exp(i\kvec\cdot\xvec)~,
\label{eq:wave}
\eneq
the configuration and the Fourier space vectors are normalized as
\beeq
\braket{\xvec}{\xvec'}=\delta^D(\xvec-\xvec')~,\quad
\braket{\kvec}{\kvec'}=(2\pi)^3\delta^D(\kvec-\kvec')~.
\eneq
Given a scalar field 
$\delta(\xvec)\equiv\braket{\xvec}{\delta}$, its three-dimensional Fourier
components are represented as
\beeq
\delta(\kvec)=\int d^3\xvec~e^{-i\kvec\cdot\xvec}~\delta(\xvec)
=\int d^3\xvec~\braket{\kvec}{\xvec}\braket{\xvec}{\delta}
=\braket{\kvec}{\delta}~,
\eneq
and its ensemble average defines the power spectrum as
\beeq
\AVE{\delta(\kvec)\delta^*(\kvec')}\equiv P(\kvec,\kvec')
=(2\pi)^3\delta^D(\kvec-\kvec')P(k)~,
\label{eq:hom}
\eneq
where the last equality holds if the power spectrum is rotationally and
translationally invariant.

With the normalization convention,
we consider a complete radial and angular basis $\ket{klm}$ in a spherical
Fourier space to decompose a scalar field in three-dimensional space.
We define its representation in configuration space as
\beeq
\braket{\bdv{x}}{klm}\equiv\sqrt{2\over\pi}~k~j_l(kr)~Y_{lm}(\Xang)~,
\eneq
where $r=|\xvec|$, $\Xang=(\theta,\phi)$ is a unit 
directional vector of~$\xvec$, $j_l(kr)$ is a spherical Bessel function,
and $Y_{lm}(\Xang)$ is a spherical harmonics.
The normalization coefficient is chosen
such that the 
basis $\ket{klm}$ is orthonormal
\bear
\braket{k'l'm'}{klm}&=&\int d^3\xvec~\braket{k'l'm'}{\xvec}\braket{\xvec}{klm}
\nonumber \\
&=&\delta^D(k-k')~\delta_{ll'}\delta_{mm'}~,
\enar
where we used
\beeq
\int_0^\infty dr~r^2~j_l(ar)~j_l(br)={\pi\over2ab}~\delta^D(a-b)~.
\eneq
By expanding the plane wave in Eq.~(\ref{eq:wave})
\bear
\braket{\xvec}{\kvec}&=& 
4\pi\sum_{lm}i^lj_l(kr)Y^*_{lm}(\Kang)Y_{lm}(\Xang) \nonumber\\
&=&(2\pi)^{3/2}\sum_{lm}{i^l\over k}Y^*_{lm}(\Kang)\braket{\xvec}{klm}~,
\label{eq:rayleigh}
\enar
and using the completeness condition of the $\ket{klm}$-basis
\beeq
\braket{\xvec}{\kvec}=
\int dk'\sum_{lm}\braket{\xvec}{k'lm}\braket{k'lm}{\kvec}~,
\eneq
we derive the relation between our spherical Fourier basis and the usual
Fourier mode
\beeq
\braket{\kvec}{k'lm}=(2\pi)^{3/2}{(-i)^l\over k}Y_{lm}(\Kang)~\delta^D(k-k')~.
\eneq
Naturally,
the spherical Fourier basis $\ket{klm}$ encodes the amplitude $k=|\kvec|$
of the three-dimensional Fourier mode $\ket{\kvec}$
and its angular direction $Y_{lm}(\Kang)$.

Based on the $\ket{klm}$-basis, any scalar field in configuration space
can be spherically decomposed as
\bear
\delta(\xvec)&=&\braket{\xvec}{\delta}=\int_0^\infty dk\sum_{lm}
\braket{\xvec}{klm}\braket{klm}{\delta} \\
&=&\int_0^\infty dk\sum_{lm}\sqrt{2\over\pi}~k~j_l(kr)~Y_{lm}(\Xang)~
\delta_{lm}(k)~, \nonumber
\enar
and its spherical Fourier mode $\delta_{lm}(k)$
is related to the three-dimensional Fourier component as
\beeq
\delta_{lm}(k)\equiv\braket{klm}{\delta}
={i^lk\over(2\pi)^{3/2}}\int d^2\Kang~Y_{lm}^*(\Kang)~\delta(\kvec)~.
\label{eq:klm}
\eneq
Due to our normalization convention of the spherical Fourier basis, the
spherical Fourier mode $\delta_{lm}(k)$ is slightly different from the
usual coefficient of the angular decomposition, which is just the angular
integral over $\Kang$. Finally, the spherical power spectrum $\SP_l(k,k')$ is 
defined as
\bear
\label{eq:sspow}
&&\AVE{\delta_{lm}(k)~\delta_{l'm'}^*(k')}
\equiv\delta_{ll'}\delta_{mm'}\SP_l(k,k') \\
&&\hspace{-10pt}
={i^l(-i)^{l'}kk'\over(2\pi)^3}
\int d^2\Kang~ d^2\Kang'~Y_{lm}^*(\Kang)Y_{l'm'}(\Kang') 
\AVE{\delta(\kvec)\delta^*(\kvec')}~. \nonumber 
\enar
For a rotationally and translationally invariant power spectrum in 
Eq.~(\ref{eq:hom}), the spherical power spectrum is
\beeq
\SP_l(k,k')=\delta^D(k-k')\SP_l(k)=\delta^D(k-k')P(k)~,
\label{eq:ddd}
\eneq
and it reduces to the three-dimensional power spectrum $\SP_l(k)=P(k)$.
Equation~(\ref{eq:ddd}) defines the three-dimensional power spectrum
$\SP_l(k)$. The three-dimensional power spectrum $\SP_l(k)$
is independent of angular
multipole~$l$, because the underlying power spectrum is isotropic.
In case of anisotropic power spectrum $P(\kvec)$, 
$\SP_l(k)$ will depend on angular
multipole~$l$.
Using the representation $\delta(\xvec)$ in configuration space,
a similar but more convenient formula can be derived for the spherical Fourier
mode as
\beeq
\label{eq:klmkey}
\delta_{lm}(k)=\int d^3\xvec~\sqrt{2\over\pi}~k~j_l(kr)~Y^*_{lm}(\Xang)
~\delta(\xvec)~,
\eneq
and its spherical power spectrum is then
\bear
\label{eq:spow}
\SP_l(k,k')&=&{2kk'\over\pi}\int d^3\xvec_1\int d^3\xvec_2~Y_{lm}^*(\Xang_1)
Y_{lm}(\Xang_2) \nonumber \\
&&\times
j_l(kr_1)j_{l}(k'r_2)~\AVE{\delta(\xvec_1)\delta(\xvec_2)}~.
\enar
Fourier components and its power spectra are dimensionful:
\bear
\left[\delta_{lm}(k)\right]&=&L^2~,\\
\left[\delta(\kvec)\right]&=&[P(\kvec)]=[\SP_l(k)]=L^3~,\nonumber \\
\left[\SP_l(k,k')\right]&=&L^4~,\quad
\left[P(\kvec,\kvec')\right]=L^6~. \nonumber
\enar

Despite the similarity to the angular power spectrum $C_l$ analysis, the
spherical Fourier power spectrum $\SP_l(k)$ differs in a key  aspect: 
the spherical Fourier analysis is three-dimensional,
utilizing information on the radial positions of galaxies, 
while two-dimensional analysis like those in the CMB literature 
lacks the radial information.
This critical difference is the advantage in the spherical Fourier analysis,
where full three-dimensional Fourier modes can be mapped, whereas in 
two-dimensional analysis radial modes are projected along the line-of-sight 
direction, contributing to the power in different angular multipoles.

\subsection{Redshift Distribution and Survey Window Function}
\label{ssec:survey}

In observation, we can only measure galaxies in the past light cone. 
Therefore, radial coordinates parametrized by the observed redshift 
carry special information, namely, {\it time} --- quantities at two
different radial coordinates are at two different redshifts. Furthermore,
the mean number density $\bar n_g(z)$ of galaxies evolves 
in time, such 
that their fluctuation field $\delta_g$ should be properly weighted by  
their redshift distribution $d\bar n_g/dz$.
This complication can be readily handled by decomposing the number density
field $n_g(\xvec)$, instead of its fluctuation field 
$\delta_g(\xvec)=n_g(\xvec)/\bar n_g(z)-1$,
where the three-dimensional position vector~$\xvec$
is understood as a function of its radial position $r(z)=|\xvec|$ and angular 
position $\Xang$.

Since the mean $\bar n_g(z)$ is independent of angular 
position,\footnote{By definition, the mean $\bar n_g(z)$ is a quantity in
a homogeneous universe, and the residual part $\delta_g(z,\Xang)$ is a
fluctuation around the mean. In observation, the mean is independent of angle 
by construction. The two means should agree, provided that $n_g(\xvec)$ is 
averaged over a large volume at a given redshift.}
it only contributes to the monopole $\delta_{00}(k)$ in Eq.~(\ref{eq:klmkey}),
and the monopole vanishes by definition at $k=0$. 
Without loss of generality, we can express $\bar n_g(z)$ as
\beeq
\bar n_g(z) \equiv \tilde n_g~\RW(z)\equiv
 \left(\frac{N_\up{tot}}{V_s}\right) \RW(z)~,
\label{eq:www}
\eneq
where the surveyed volume is
\beeq
V_s=4\pi\int_0^\infty dr~r^2\RW=4\pi\int_0^\infty dz~{r^2\over H}~\RW~,
\label{eq:vol}
\eneq
and $N_\up{tot}$  is the total number of galaxies measured in the survey. 
Equation~(\ref{eq:www}) 
defines the survey window function $\RW$, also known as radial selection
function. It is related to the redshift distribution as
\beeq
\PP(z)={r^2\over H}~\RR(r)={4\pi\over V_s}{r^2\over H}~\RW(z)
={4\pi\over N_\up{tot}}{r^2\over H}~\bar n_g(z)~,
\label{eq:zrel}
\eneq
where $H(z)$ is the Hubble parameter and the normalization convention is 
$1=\int dz~\PP(z)=\int dr~r^2~\RR(r)$. In principle, $\RW(z)$ could be 
generalized to include an angular selection function. In what follows 
however, we will consider galaxy surveys with full-sky coverage
and uniform angular selection function for simplicity. 
The galaxy number density at position $\xvec$ thus is
\beeq
n_g(\xvec) = \tilde n_g \RW(z)\left[1+\delta_g(z,\Xang)\right]~,
\label{eq:ngx}
\eneq
and its two-point correlation function reads
\bear
\label{eq:2ptngx}
\AVE{n_g(\xvec_1)n_g(\xvec_2)} &=& \tilde n_g^2\RW(z_1)\RW(z_2)
\Bigl[1+\xi_g(\xvec_2-\xvec_1)\Bigr] \nonumber \\
&& +\tilde n_g \RW(z_1)\delta^D(\xvec_2-\xvec_1)~,
\enar
where $\xi_g(\xvec_2-\xvec_1)$ 
is the Fourier transform of the noise-free galaxy power spectrum.
We have assumed that the galaxies are an (inhomogeneous) Poisson sampling of
$1+\delta_g(\xvec)$ to derive the shot-noise term \cite{FEKAPE94}.

Before we proceed further, we introduce the transfer functions $\TT_X(k,z)$
of random perturbation variables~$X$ that further simplify the spherical
Fourier decomposition by separating radial (time) dependence from angular 
dependence. For the galaxy fluctuation, we have
\beeq
\delta_g(\kvec,z)=\TT_g(k,z)~\cp(\kvec) + \epsilon(\kvec,z)~,
\eneq 
where the power spectrum of the comoving curvature 
$\Delta^2_{\cp}\!(k)=k^3P_{\cp}(k)/2\pi^2$ 
at initial epoch is a nearly 
scale-invariant and the transfer function is independent of angle.
The comoving curvature $\cp(\kvec)$
is often denoted as~$\zeta(\kvec)$ in literature. Arising from the discrete
distribution, 
$\epsilon(\kvec,z)$ is a residual Poisson noise that is uncorrelated with
$\delta_g(\kvec,z)$, with power spectrum 
$\AVE{\epsilon(\kvec,z)\epsilon(\kvec',z)}=(2\pi)^3\delta^D(\kvec-\kvec')
/(\tilde n_g\RW(z))$.

In case that time evolution is related to radial coordinates,
it is more natural to use Eq.~(\ref{eq:klmkey}) than Eq.~(\ref{eq:klm}), and
so is it to use Eq.~(\ref{eq:spow}) than Eq.~(\ref{eq:sspow}) for computing 
spherical Fourier modes and their spherical power spectrum, respectively.
On inserting Eq.~(\ref{eq:ngx}) into Eq.~(\ref{eq:klmkey}) and substituting 
the transfer function~$\TT_g$ 
and survey selection function~$\RW$, the spherical Fourier
mode simplifies to
\bear
\delta_{lm}(k) &=& i^l\int\!\!{d\ln k'k'^3\over2\pi^2}\int\!\! d^2\Kang'~
\cp(\kvec')~Y_{lm}^*(\Kang')~\ST_l(k',k) \nonumber \\
&& +~ \epsilon_{lm}(k)~, 
\label{eq:sfm}
\enar
where $\epsilon_{lm}(k)$ is the spherical Fourier transform of the residual
noise field $\tilde n_g\RW(z) \epsilon(\xvec)$, and the spherical multipole 
function $\ST_l(k',k)$ is defined as
\beeq
\ST_l(\tilde k,k)\equiv k~\sqrt{2\over\pi}
\int_0^\infty dr~r^2~\RW(r)~j_l(\tilde kr)~j_l(kr)~\TT_g(\tilde k,r)~,
\label{eq:sfmm}
\eneq
where 
its dimension is $[\ST_l(\tilde k, k)]=L^2$. After some simplification, the 
spherical power spectrum in Eq.~(\ref{eq:spow}) eventually reads
\bear
\SP_l(k,k') &=& 4\pi \tilde n_g^2
\int\!\! d\ln\tilde k~\Delta^2_{\cp}\!(\tilde k) \ST_l(\tilde k,k) 
\ST_l(\tilde k,k') \nonumber \\
&& + \frac{2 k k'}{\pi} \tilde n_g
\int_0^\infty\!\!dr~ r^2 \RW(r) j_l(kr) j_l(k' r)~.
\label{eq:sfp}
\enar
The second-term in the right-hand side is the shot-noise contribution. Using
the Limber approximation (see Sec. \ref{ssec:limber}),
the shot-noise power spectrum can be rewritten as 
\bear
\NP_l(k,k') &\equiv& \frac{2 k k'}{\pi} \tilde n_g
\int_0^\infty\!\!dr~ r^2 \RW(r) j_l(kr) j_l(k' r) \nonumber \\
&\approx& \tilde n_g \RW(\nu/k) \delta^D(k-k')~,
\enar
where $\nu=l+1/2$. For a homogeneous and isotropic galaxy population 
with constant comoving number density $\bar n_g=\tilde n_g$ and power spectrum 
$\AVE{\delta_g(\kvec)\delta_g^*(\kvec')}=(2\pi)^3\delta^D(\kvec-\kvec')P_g(k)$,
the spherical power spectrum Eq.~(\ref{eq:sfp}) yields the well-known relation
\beeq
\SP_l(k) = \bar n_g^2 P_g(k)+\bar n_g~.
\label{eq:hisfp}
\eneq
The angular multipole~$l$ controls the transverse wavenumber $k_\perp$ 
but, since the amplitude of the wavevector~$k=|\kvec|$ is already set 
in $\SP_l(k)$, the spherical power spectrum must be independent of~$l$.
In practice, however, $\RW(r)$ is always different than unity so that 
$\SP_l(k,k)$ is never independent of the multipole $l$. Note also that, with
the galaxy number density defined as Eq.~(\ref{eq:ngx}), both the
spherical power spectra $\SP_l$ and $\NP_l$ have dimensions of $L^{-2}$.
We will henceforth assume that $V_s$ is accurately known and work 
with the normalized galaxy number density $n_g(\xvec)/\tilde n_g$, instead of 
$n_g (\xvec)$. 
In this case, the spherical power spectrum is given by the right-hand side 
of Eq.~(\ref{eq:sfp}) divided by $\tilde n_g^2$.

Before we close this section and apply the spherical Fourier analysis 
to the general relativistic description of galaxy clustering, we discuss
our assumption for survey geometry and other approaches to handling the
survey window function.
Our spherical Fourier decomposition assumes that the full sky 
is available for measuring the galaxy number density field $n_g(\xvec)$.
For surveys with an incomplete sky coverage, angular multipoles 
with characteristic scale
larger than the sky coverage are simply unavailable, while angular multipoles
on smaller scales remain unaffected.

In literature, the radial boundary condition is often imposed by choosing
discrete wavenumbers~$k_{n_l}$, in which the range of integer~$n_l$ depends 
on angular multipole~$l$ (e.g., \cite{FILAET95,HETA95,SHCRPE12}).
Since the galaxy number density can be
measured only within the survey region, the number density field 
outside the survey volume is set zero in those approaches.
Hence, this situation is equivalent
to solving the Poisson equation within the survey area and
the Laplace equation outside the survey \cite{FILAET95}. 
A unique solution can be singled out on imposing Dirichlet or 
Neumann conditions at the boundary (or a linear combination thereof) 
\cite{HETA95}. 
Therefore, despite the same number density
measured within the survey region, the resulting potential and velocity fields
are different due to the non-locality of their relation to the density
field, depending on the choice of boundary conditions.
As we are interested in describing the observed galaxy number density 
$n^\up{obs}_g(\xvec)$,
rather than the other derived quantities such as velocity or potential, 
no boundary condition need be
imposed, and the results for $n^\up{obs}_g(\xvec)$
are identical in all approaches, whenever
our continuous wavenumbers equal discrete ones or the survey volume becomes
infinite.

\section{Galaxy Clustering in General Relativity}
\label{sec:gcgr}
Here we present the full relativistic description of galaxy clustering
in Sec.~\ref{ssec:gr}. Weight functions are derived to facilitate the full
calculations of the spherical galaxy power spectrum in Sec.~\ref{ssec:weight}
and simplified Limber formulas are presented in Sec.~\ref{ssec:limber}
to provide physical insight.

\subsection{General Relativistic Description of Galaxy Clustering}
\label{ssec:gr}
Since the standard approach to modeling galaxy clustering is based on the 
Newtonian framework, it naturally breaks down on large scales, 
where relativistic effect becomes significant. 
Recently, the fully relativistic description of galaxy clustering was formulated 
\cite{YOFIZA09,YOO10}, resolving gauge issues between observable and unobservable 
quantities.
The key finding in the theoretical development is that unobservable quantities
such as the true redshift and the unlensed 
galaxy position are gauge-dependent. 
Therefore, we need to construct galaxy clustering statistics based
on observables quantities like the galaxy redshift and the position on the sky.
In doing so, effects such as  redshift-space distortions and 
lensing magnification are naturally incorporated. 
Hence, this approach provides a unified description of various 
effects in galaxy clustering \cite{YOO09}.

The relativistic description of galaxy clustering can be
derived from the fact
that the number $dN_g^\up{obs}$ of observed galaxies in a small volume
is conserved:
\beeq
dN_g^\up{obs}=n^\up{obs}_g dV_\up{obs}=n^\up{phy}_g dV_\up{phy}~,
\label{eq:conserv}
\eneq
and this equation defines the observed galaxy number density $n_g^\up{obs}$
with the observed volume element
\beeq
dV_\up{obs}(z)={r^2(z)\over H(z)}~\sin\theta ~dz~d\theta ~d\phi~,
\label{eq:cvol}
\eneq
described by the observed galaxy position $(\theta,\phi)$ on the sky and 
the observed redshift~$z$. The ``observed'' volume element $dV_\up{obs}$
is different from the ``physical'' volume $dV_\up{phy}$ occupied by the
observed galaxies on the sky, and the distortion between these two volume
elements, so called the volume effect, 
gives rise to contributions to the observed galaxy fluctuation. While the 
dominant contribution in the volume effect is the
redshift-space distortion and the gravitational lensing effect, there
exist other subtle relativistic contributions, and the physical volume
can be obtained by tracing the photon path backward  (see \cite{YOFIZA09,YOO10}
for details in deriving $dV_\up{phy}$ and computing the volume effect).

Another source of fluctuations in galaxy clustering is the 
source effect describing the distortion associated with the physical galaxy
number density $n_g^\up{phy}$. 
The mean galaxy number density is obtained by averaging the
observed galaxy number density at the observed redshift, which differs
from the proper time at the galaxy rest frame. 
The contribution of this source effect is proportional to the evolution bias 
factor of the galaxy number density 
\beeq
e=3+{d\ln\bar n_g\over d\ln(1+z)}~,
\label{eq:ebias}
\eneq
where the factor three in Eq.~(\ref{eq:ebias})
appears because we use the comoving number density $\bar n_g$.
Other contributions of the source effect can arise, depending on how we
define the galaxy sample at hand (see \cite{YOO09,YOFIZA09,YOO10,YOHAET12}).
In typical galaxy surveys, galaxy samples are defined with 
the observed luminosity inferred from the observed flux, 
which is different from the intrinsic luminosity. So the source effect 
associated with it is then proportional to the luminosity function slope
\beeq
p=-0.4~{d\log\bar n_g\over d\log L}~,
\eneq
where the luminosity~$L$ is computed by using the observed flux and
the observed redshift.

Putting all these effects together and accounting for the relativistic
contributions, the general relativistic description of the observed galaxy 
fluctuation can be written, to the linear order in perturbation, as
\cite{YOFIZA09,YOO10} 
\bear
\delta^\up{obs}_g&=&b~\dv-e~\dz_v-5p~\ddL+\ax+2~\px+V\nonumber \\
&&+3~\dzg-H{d\over dz}\left({\dzg\over\HH}\right)+2~{\drg\over r} -2~\kag~,
\label{eq:fullGR}
\enar
where the linear bias factor is~$b$, the comoving gauge matter density
is $\dv$, the lapse term $\dz$ in the observed redshift is defined as
$1+z=(1+\bar z)(1+\dz)$,
the dimensionless fluctuation in the luminosity distance is $\ddL$, 
the gauge-invariant temporal and spatial 
metric perturbations are $\ax$ and $\px$, the line-of-sight velocity is~$V$, 
and the gauge-invariant radial displacement and lensing convergence are $\drg$ 
and $\kag$.
The subscripts~$\chi$ and~$v$ represent that 
the gauge-invariant quantities with the corresponding subscript are identical
to those quantities evaluated in the conformal Newtonian gauge ($\chi=0$)
or in the dark matter comoving gauge ($v=0$), 
in which the shear seen by the normal observer vanishes ($\chi=0$) or
the off-diagonal component of the energy momentum tensor vanishes ($v=0$),
respectively. Note that $\ax$ and $\px$ correspond to the Bardeen's
variables $\Phi_A$ and $\Phi_H$ \cite{BARDE80} and
we have assumed no vector and tensor contributions in Eq.~(\ref{eq:fullGR}).
We refer the reader to \cite{YOFIZA09,YOO10,YOHAET12} 
for details in the derivation 
(see also \cite{CHLE11,BODU11,BASEET11,BRCRET12,JESCHI12}).

Equation~(\ref{eq:fullGR}) is written in a physically transparent way.
The observed galaxy fluctuation is modulated by the matter density $\dv$, and 
there exist additional contributions from the source and the volume effects.
The source effect is composed of $e~\dz_v$ and $5p~\ddL$, and the other
terms in Eq.~(\ref{eq:fullGR})
come from the volume effect. Of the volume effect, 
the last three terms in the first line of
Eq.~(\ref{eq:fullGR}) defines the rest frame of the observed galaxies,
and the remaining 
terms in the second line describe the distortion of the comoving volume element
in Eq.~(\ref{eq:cvol}). It is noted that the term $3~\dzg$ accounts for 
the distortion in the comoving volume factor~$a^3$ and in Eq.~(\ref{eq:cvol})
the conversion between the comoving and the physical volume elements
is based on the observed redshift.

In order to compute the observed galaxy fluctuation $\delta_g^\up{obs}$,
we need to evaluate the individual gauge-invariant variables in 
Eq.~(\ref{eq:fullGR}). While Eq.~(\ref{eq:fullGR}) is arranged in terms
of gauge-invariant variables to explicitly ensure the gauge-invariance 
of $\delta_g^\up{obs}$, it can be computed with any choice of gauge 
conditions, but preferably with the most convenient for computation.
Using the Einstein equations, we have the following relation 
for a flat $\Lambda$CDM universe \cite{BARDE80,KOSA84,HWNO01,HWNO99R,HWNO05},
\bear
\phi&\equiv&\px=-\ax={3H_0^2\over2}~{\OM\over ak^2}~\dm ~,\\
v&\equiv&\vx=-{1\over k}~\dm'=-{\HH f\over k}~\dm~,
\enar
where the comoving gauge matter density is 
obtained upon setting $\dm\equiv\dv$. and
the logarithmic growth rate is $f=d\ln\dm/d\ln a$. Other gauge-invariant
variables in Eq.~(\ref{eq:fullGR}) can be readily expressed
in terms of $\dm$, $v$, and $\phi$ as \cite{YOHAET12}
\bear
\label{eq:los}
V&=&{\partial\over \partial r}
\int{d^3\kvec\over(2\pi)^3}{-v\over k}~ 
e^{i\kvec\cdot\xvec}~, \\
\label{eq:dz}
\dzg&=&V+\phi+\int_0^r d\tilde r~2\phi'~, \\
\label{eq:dzv}
\dz_v&=&\dzg+\int{d^3\kvec\over(2\pi)^3}~{\HH v\over k}~
e^{i\kvec\cdot\xvec}~, \\
\label{eq:dr}
\drg&=&-{\dzg\over\HH}-\int_0^rd\tilde r~2\phi~,\\
\label{eq:kappa}
\kag&=&-\int_0^rd\tilde r\left({r-\tilde r\over \tilde r r}\right)
\hat\nabla^2\phi~,\\
\label{eq:dl}
\ddL&=&{\drg\over r}+\dzg+\phi-\kag~, \\
\label{eq:zdist}
&&\hspace{-40pt}
-H{d\over dz}\left({\dzg\over\HH}\right)=-V-{1+z\over H}\phi'-{1+z\over H}
{\partial V\over\partial r} \nonumber \\
&&-\dzg+{1+z\over H}{dH\over dz}~\dzg~.
\enar
We also note that the total derivative in Eq.~(\ref{eq:zdist}) is along 
the past light cone: $dr=dz/H$ and $dr= \partial r-\partial\tau$.

\begin{table*}
\caption{Weight functions of the observed galaxy fluctuation in 
Eq.~(\ref{eq:fullDecom}). All the perturbation variables in 
Eq.~(\ref{eq:fullGR}) are scaled with the matter density in the dark-matter
comoving gauge by using the Einstein equations, and its relations to the
matter density are defined as the weight functions that can be used for
computing the spherical multipole function in Eq.~(\ref{eq:MM}) and
the spherical power spectrum in Eq.~(\ref{eq:sffp}). Time dependent quantities
$a$, $H$, and $f$ are evaluated at the observed redshift~$z$, and 
those with tilde depend on the line-of-sight distance $\tilde r$ (or the
corresponding redshift $\tilde z\leq z$).}
\begin{ruledtabular}
\begin{tabular}{ccc}
gauge-invariant quantity && weight function $\WW_l(r,\tilde r, k)$ \\
\hline
$\dv$&& $\delta^D(r-\tilde r)$ \\
$\ax$ && $-\delta^D(r-\tilde r){3H_0^2\over2}{\OM\over ak^2}$ \\
$\px$ && $\delta^D(r-\tilde r){3H_0^2\over2}{\OM\over ak^2}$ \\
$V$ && $\delta^D(r-\tilde r)\left({\HH f\over k}{\partial\over k\partial r}
\right)$ \\
$\dzg$ && $\delta^D(r-\tilde r)\left[
\left({\HH f\over k}{\partial\over k\partial r}\right)+
{3H_0^2\over2}{\OM\over ak^2}\right]+3H_0^2\OM{\tilde H(\tilde f-1)\over 
k^2}$ \\
$\dz_v$ &&  $\delta^D(r-\tilde r)\left[
\left({\HH f\over k}{\partial\over k\partial r}\right)+
{3H_0^2\over2}{\OM\over ak^2}-{\HH^2f\over k^2}\right]+3H_0^2\OM{\tilde H
(\tilde f-1)\over k^2}$ \\
${\drg\over r}$ && $-{\delta^D(r-\tilde r)\over\HH r}\left[
\left({\HH f\over k}{\partial\over k\partial r}\right)+
{3H_0^2\over2}{\OM\over ak^2}\right]-{3H_0^2\over\HH r}~\OM
{\tilde H(\tilde f-1)\over k^2} -{3H_0^2\over r}{\OM\over\tilde a k^2}$\\
$-H{d\over dz}\left({\dzg\over\HH}\right)$ &&  
$-\delta^D(r-\tilde r)\left[\left(2-{1+z\over H}{dH\over dz}\right)
\left({\HH f\over k}{\partial\over k\partial r}
\right) +{3H_0^2\over2}{\OM\over ak^2}\left(f-{1+z\over H}{dH\over dz}\right)
+f{\partial^2\over(k\partial r)^2}\right]
-3H_0^2\OM\left(1-{1+z\over H}{dH\over dz}\right)
{\tilde H(\tilde f-1)\over k^2}$ \\
$\kag$ && ${3H_0^2\over2}\OM~l(l+1)\left({r-\tilde r\over\tilde r r}\right)
{1\over \tilde a k^2}$ \\
$\ddL$ && $\delta^D(r-\tilde r)\left[\left(1-{1\over\HH r}\right)
\left({\HH f\over k}{\partial\over k\partial r}\right)+
{3H_0^2\over2}{\OM\over ak^2}\left(2-{1\over\HH r}\right)\right]
+{3H_0^2\OM\over k^2}\left[\tilde H(\tilde f-1)\left(1-{1\over\HH r}\right)
-{1\over\tilde a r}-{l(l+1)\over2~\tilde a}
\left({r-\tilde r\over\tilde r r}\right)\right]$ \\
\end{tabular}
\end{ruledtabular}
\label{tab:weight}
\end{table*}

\subsection{Fourier Decomposition and Weight Function}
\label{ssec:weight}
Now that all the contributions in Eq.~(\ref{eq:fullGR}) can be expressed
in terms of the matter density~$\dm$ and its transfer function $\TTm(k,r)$,
we define the weight functions $\WW_i(\xvec,\kvec)$ for each
contribution in Eq.~(\ref{eq:fullGR}) through the following 
decomposition of the observed galaxy fluctuation:
\bear
\hspace{-5pt}
\delta_g^\up{obs}(\xvec) &=& \sum_i\int_0^r\!\! d\tilde r\int\!\!{d^3\kvec\over(2\pi)^3}~
\WW_i(\tilde\xvec,\kvec)~\TTm(k,\tilde r)~\cp(\kvec)~
e^{i\kvec\cdot\tilde\xvec} \nonumber \\
&& + ~ \epsilon(\xvec)~,
\label{eq:fullDecom}
\enar
where the observed galaxy position is $\xvec=[r(z),\Xang]$ and 
the line-of-sight position is $\tilde\xvec=(\tilde r,\Xang)$ 
in spherical coordinates ($\xvec/\!/\tilde\xvec$). The dimension of the weight
functions is $[\WW_i(\tilde\xvec,\kvec)]=L^{-1}$, and
$\epsilon(\xvec)$ is the residual shot-noise field.

Categorically, each contribution in Eq.~(\ref{eq:fullGR}) can be classified in
three different types according to its dependence on the angle
$\Xang$ and the line-of-sight distance $\tilde r$. 
The first and simplest type includes all the contributions
such as $\dv$, $\ax$, and $\px$ that are independent of $\Xang$ and~$\tilde r$. 
For example, the matter density in Eq.~(\ref{eq:fullGR}) is
\beeq
\dv(\xvec)=\dm(\xvec)=\int{d^3\kvec\over(2\pi)^3}~\TTm(k, r)~\cp(\kvec)~
e^{i\kvec\cdot\xvec}~,
\eneq
and hence the weight function for the matter density takes the simplest form
\beeq
\WW(\tilde\xvec,\kvec)=\delta^D(r-\tilde r)~.
\eneq
Similarly, the weight function for the potential $\px=\phi$ is
\beeq
\WW(\tilde\xvec,\kvec)=\delta^D(r-\tilde r)~{3H_0^2\over2}{\OM\over ak^2}~.
\eneq

Fluctuations of the second type depend on the observed angle, 
yet are independent of the line-of-sight distance. 
This applies to the line-of-sight velocity contribution~$V$ (and functions 
thereof such as $\dzg$, $\drg$, $\ddL$),
\bear
\label{eq:VVV}
V(\xvec)&=&{\partial\over\partial r}\int{d^3\kvec\over(2\pi)^3}
\left(\HH f{\dm\over k^2}\right)~e^{i\kvec\cdot\xvec} \\
&=&\int{d^3\kvec\over(2\pi)^3}{\HH f\over k}\TTm(k,r)~\cp(\kvec)
\left({\partial\over k\partial r}\right)e^{i\kvec\cdot\xvec}~, \nonumber
\enar
of which the weight function is
\beeq
\WW(\tilde\xvec,\kvec)=\delta^D(r-\tilde r)
\left({\HH f\over k} {\partial\over k\partial r}\right)~.
\eneq
The derivative operator acts on the radial component of 
the plane wave in Eq.~(\ref{eq:fullDecom}).

Finally, the third type covers the contributions along the 
line-of-sight direction that are independent of the observed angle. 
This category includes numerous line-of-sight integrals 
in Eqs.~(\ref{eq:dz})$-$(\ref{eq:dl}), including the Sachs-Wolfe effect and 
the weak lensing convergence. 
For example, the Sachs-Wolfe contribution in $\dzg$ in Eq.~(\ref{eq:dz}) 
can be decomposed as
\bear
&&\hspace{-10pt}
\int_0^r d\tilde r~\phi'=\int_0^r d\tilde r\int{d^3\kvec\over(2\pi)^3}
~\phi(\kvec,\tilde r)'
e^{i\kvec\cdot\tilde\xvec}  \\
&&\hspace{-10pt}
=\int_0^r d\tilde r\int{d^3\kvec\over(2\pi)^3}
\left[{3H_0^2\over2}\OM{\tilde H(\tilde f-1)\over k^2}\right]
\TTm(k,\tilde r)~\cp(\kvec)~e^{i\kvec\cdot\tilde\xvec} ~, \nonumber
\enar
and, therefore, the resulting weight function is 
\beeq
\WW(\tilde\xvec,\kvec)={3H_0^2\over2}\OM{\tilde H(\tilde f-1)\over k^2}~,
\eneq
where $\tilde H$ and $\tilde f$ depend on $\tilde r$ and we used 
$\phi'=\HH\phi(f-1)$.
Another example of this type is the weak lensing convergence in 
Eq.~(\ref{eq:kappa})
\bear
\kag&=&-\int_0^rd\tilde r\left({r-\tilde r\over \tilde r r}\right)
\hat\nabla^2\int{d^3\kvec\over(2\pi)^3}~\phi(\kvec,\tilde r)~
e^{i\kvec\cdot\tilde\xvec} \nonumber \\
&=&-\int_0^rd\tilde r\int{d^3\kvec\over(2\pi)^3}
\left[\left({r-\tilde r\over \tilde r r}\right)
{3H_0^2\over2\tilde ak^2}\OM\right] \nonumber \\
&&\times
\TTm(k,\tilde r)~\cp(\kvec)~\hat\nabla^2 e^{i\kvec\cdot\tilde\xvec} ~.
\enar
The weight functions depend on position~$r$, wavevector~$k$, and line-of-sight
distance~$\tilde r$, and its angular dependence can be 
removed by using the plane  wave expansion, i.e.,
\beeq
\WW_l(r,\tilde r,k)={3H_0^2\over2}\OM~
l~(l+1)\left({r-\tilde r\over\tilde rr}\right){1\over \tilde ak^2}~,
\eneq
where the expansion factor~$\tilde a$ depends on $\tilde r$ and
the angular part of the plane wave satisfies
$\hat\nabla^2Y_{lm}(\Xang)=-l(l+1)Y_{lm}(\Xang)$.
For convenience, Table~\ref{tab:weight} summarizes all the 
weight functions.

With the full Fourier decomposition of the observed galaxy fluctuation,
the spherical multipole function in Eq.~(\ref{eq:sfmm}) must 
be generalized as follows to take into proper account each component 
in Eq.~(\ref{eq:fullDecom}),
\bear
\label{eq:MM}
\ST^i_l(\tilde k,k)&\equiv&k\sqrt{2\over\pi}
\int_0^\infty dr~r^2~\RW(r)~j_l(kr)  \\
&&\times\int_0^rd\tilde r~\WW_l^i(r,\tilde r,\tilde k)
~\TTm(\tilde k,\tilde r)~j_l(\tilde k\tilde r)~ ,\nonumber
\enar
and the spherical multipole function of the observed galaxy fluctuation is 
simply
\beeq
\ST_l(\tilde k,k)=\sum_i\ST^i_l(\tilde k,k)~.
\eneq
The spherical power spectrum of the observed (normalized)
galaxy fluctuation is then 
\beeq
\SP_l(k,k') \equiv \bar{\SP}_l(k,k') + \NP_l(k,k') ~,
\eneq
where $\bar{\SP}_l(k,k')$ is the noise-free galaxy power spectrum,
\beeq
\bar{\SP}_l(k,k') = 4\pi
\int d\ln\tilde k~\Delta^2_{\cp}\!(\tilde k)~\ST_l(\tilde k,k)~
\ST_l(\tilde k,k') ~,
\label{eq:sffp}
\eneq
and the noise component is
\beeq
{\cal N}_l(k,k')\equiv \frac{2 k k'}{\pi \tilde n_g}
\int_0^\infty\!\!dr~ r^2 \RW(r) j_l(kr) j_l(k' r)~,
\eneq
where $\SP_l$ is identical to Eq.~(\ref{eq:sfp}) except 
for a factor of $\tilde n_g^2$.
Its dimension thus is $L^4$.

The weight functions defined here
differ from those in \cite{YOFIZA09}, as we scale all the perturbation 
variables with the matter density in the comoving gauge. This approach is
better suited to numerical computation, but its applicability 
is limited to a flat universe with a pressureless component, including a 
cosmological constant and baryons on large scales. 
For different cosmological models or modified gravity theories, the relation 
among $\dv$, $\ax$, $\px$, and $\vx$ is in general different 
\cite{LOMERI12,HABOCH12,LOYOKO13}
from what we 
assumed for a pressureless medium, and one should use the weight functions in 
\cite{YOFIZA09} with the corresponding transfer functions.

\subsection{Limber Approximation}
\label{ssec:limber}
Before we present numerical calculations of the spherical power spectrum
in Sec.~\ref{sec:results}, we adopt the Limber approximation to
derive analytic formulas for the spherical multipole functions and 
the spherical power spectra.
The Limber approximation (\cite{LIMBE53,KAISE92,LOAF08}) relies on the fact 
that 
the spherical Bessel function $j_l(x)$ peaks at $x\approx \nu\equiv l+1/2$
and oscillates rapidly for $x\gtrsim \nu$. For large values of angular 
multipole~$l$, 
integrals of product of a smooth function $f(x)$ times $j_l(x)$ can be
approximated by
\beeq
\int_0^\infty\!\!dx~ f(x)~ j_l(x)\simeq\sqrt{\frac{2}{\pi\nu}}f(\nu)
\left[1+{\cal O}(1/\nu^2)\right]~.
\eneq
With aid of the orthogonality relation for the spherical Bessel functions, 
this result can also be written as \cite{BEBOET11}
\bear
\frac{\pi}{2}\int_0^\infty\!\!dr~r^2 f(r) j_l(kr) j_l(k' r) &\simeq&
\frac{\delta^D(k-k')}{k^2} \\ 
&& \times f(\nu/k)\left[1+{\cal O}(1/\nu^2)\right]~. \nonumber
\nonumber
\enar

In the case of the matter density, for instance, we obtain the spherical
multipole function 
\beeq
\ST^\delta_l(\tilde k,k)\approx \sqrt{\frac{\pi}{2k^2}}~\RW(\nu/k)
\TTm\left(k,\frac{\nu}{k}\right) \delta^D(k-\tilde k)~,
\eneq
and the spherical power spectrum
\beeq
\SP^\delta_l(k,k') \approx \RW(\nu/k)^2 P_m\!\left(k,\frac{\nu}{k}\right)
\delta^D(k-k')~.
\eneq
Similarly, on using the Limber approximation,
the spherical power spectrum $\SP^\phi_l(k,k')$ of the potential perturbation
can be readily computed owing to the simple dependence of the
spherical multipole function on the weight function.
The noise contribution to the spherical power spectrum falls in the same
category, and the Limber approximation gives
\beeq
\NP_l(k,k')\approx \frac{1}{\tilde n_g}\RW(\nu/k)~\delta^D(k-k')~.
\eneq

In addition, the Limber approximation can greatly simplify the computation
of the projected quantities such as the Sachs-Wolfe contribution and the 
gravitational lensing contribution. These quantities contain two 
line-of-sight integrations in the spherical multipole function in 
Eq.~(\ref{eq:MM}), each of which involves the spherical Bessel function.
With the Limber approximation, 
the spherical multipole function of the gravitational lensing is 
\bear
\ST^{\kag}_l(\tilde k, k)&\simeq&
\sqrt{\pi\over2\tilde k^2}~\RW(\nu/k)~{3H_0^2\over2}~\OM~
\TTm(\tilde k,\nu/\tilde k) \nonumber \\
&&\times l(l+1)\left({\tilde k-k\over \tilde ak^2\tilde k^2}\right)~,
\enar
where $\tilde a$ is the expansion factor at $\tilde r=\nu/\tilde k$. The
spherical power spectrum $\SP^\kag_l(k,k')$
of the gravitational lensing is then obtained
by integrating the spherical multipole function over $\tilde k$, instead
of a quintuple integration over $\tilde k$ and two pairs $(r, \tilde r)$. 

The spherical multipole function of the line-of-sight velocity or
 the redshift-space distortion involves the derivatives of the spherical
Bessel function, since their weight function acts on the radial component
as in Eq.~(\ref{eq:VVV}). Since derivatives of the spherical 
Bessel function $j_l(x)$ are linear combinations of spherical Bessel functions 
at different angular multipoles, they do not form an orthogonal basis.
In these cases, we find that the Limber approximation becomes less accurate,
and next-leading corrections are required to enhance the accuracy. 
Despite this shortcoming in a few cases, the 
Limber approximation provides nonetheless a simple and physical explanation 
of the complicated spherical power spectrum. 
Therefore, we will frequently rely on it when we discuss the physical
interpretation of 
the full numerical results in Sec.~\ref{sec:results}.

\begin{figure}
\centerline{\psfig{file=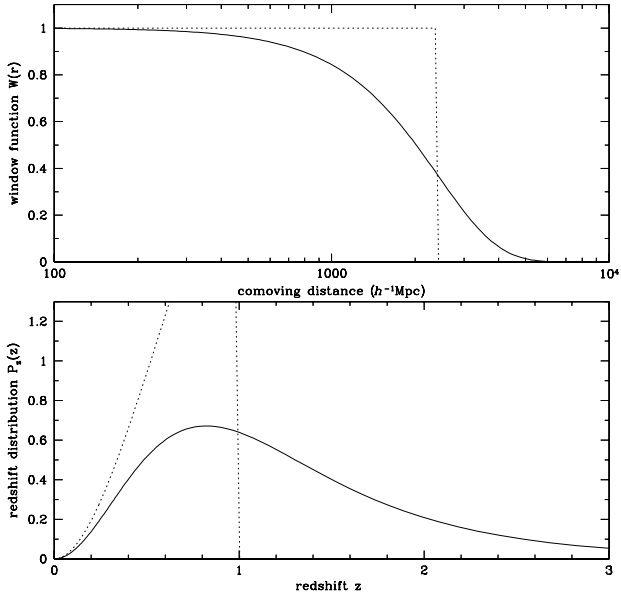, width=3.4in}}
\caption{Survey window functions $\RW$ and redshift distributions $\PZ$.
Upper panel: Gaussian window function (solid) in Eq.~(\ref{eq:gauss})
and top-hat window function (dotted) in Eq.~(\ref{eq:tophat}).
Bottom panel: Corresponding redshift distributions obtained by using 
Eq.~(\ref{eq:zrel}). The characteristic scales $r_0$ of the surveys are set
equal to the comoving distance to $z=1$. The comoving number density is 
related to the survey window function as in Eq.~(\ref{eq:www}).}
\label{fig:window}
\end{figure}

\section{Observed Spherical Power Spectrum}
\label{sec:results}
Here we present the observed spherical power spectrum in 
Sec.~\ref{ssec:spow} and compare to the corresponding flat-sky 
three-dimensional power spectrum.
The measurement uncertainties associated with the spherical power spectrum
are presented in Sec.~\ref{ssec:uncer}.

\begin{figure*}
\centerline{\psfig{file=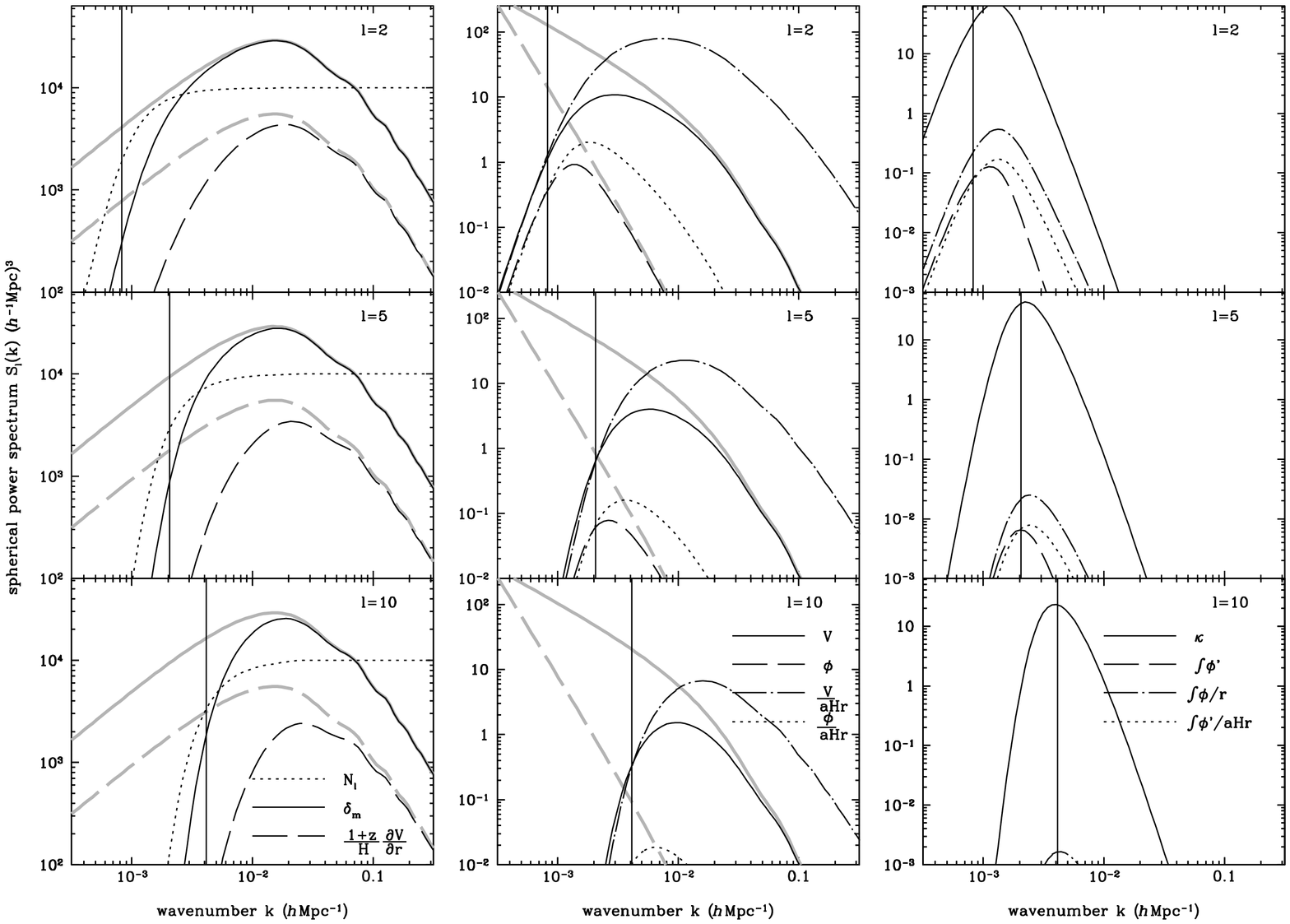, width=5.0in, angle=-90}}
\caption{Spherical power spectra with Gaussian window function. In each panel,
various curves with labels show the spherical power spectra of representative 
components of the full general relativistic formula $\delta_g^\up{obs}$ in 
Eq.~(\ref{eq:fullGR}), which are further decomposed in 
Eqs.~(\ref{eq:los})$-$(\ref{eq:zdist}).
The time evolution of the transfer and the weight functions is ignored, and
their values are set equal to those at $z=0$.
Each column plots the same spherical power spectra at angular multipoles,
ranging from $l=2$ (top rows) to $l=10$ (bottom rows),
illustrating their angular dependence. 
Vertical lines mark the characteristic
wavenumber $k_c\equiv l/r_0$, below which spherical power spectra are 
suppressed due to the constraint $k\geq k_\perp\simeq k_c$. Gray curves
show the three-dimensional power spectra at $z=0$, if there exist well-defined 
correspondences. For example, spherical power spectra for the density (solid)
and the redshift-space distortion (dashed) in the first column reduce
to the corresponding three-dimensional power spectra on small scales.
The noise power spectrum (dotted) is shown in the first column, assuming
$\tilde n_g=10^{-4}~(\hmpc)^{-3}$. Compared to the contributions in
the first two columns, the projected quantities in the third column are
negligible, except the gravitational lensing convergence.
The turn-over of the spherical power spectra on large scales reflects the
limit set by survey depth $\sim$ a few times $1/r_0$. These large-scale modes
can be recovered by deconvolving the survey window function.}
\label{fig:gaussian}
\end{figure*}

\subsection{Spherical Power Spectrum}
\label{ssec:spow}
For definiteness we consider all-sky galaxy surveys with two different 
survey window functions. 
Figure~\ref{fig:window} plots two survey window functions~$\RW(r)$ and their
corresponding redshift distribution~$\PP(z)$. 
We consider two specific window functions that approximately represent
a flux-limited survey (solid) and a volume-limited survey (dotted), 
if the comoving number
density $\bar n_g(z)$ is constant. Survey window functions (top panel)
are related to galaxy redshift distributions (bottom panel)
via Eqs.~(\ref{eq:www}) and~(\ref{eq:zrel}),
and hence to the evolution bias factor in Eq.~(\ref{eq:ebias}).
The former window function is modeled as a Gaussian (solid)
\beeq
\RW(r)=\exp\left[-\left({r\over r_0}\right)^2\right]~,
\label{eq:gauss}
\eneq
and the latter window function is modeled as a top-hat (dotted)
\beeq
\RW(r)=\bigg\{
\begin{array}{ccc}1~,&& r\leq r_0~,\\0~, && r>r_0~,\end{array}
\label{eq:tophat}
\eneq
where the characteristic distance or the survey depth is
$r_0=2354\hmpc$ set equal to the comoving distance to $z=1$.
In both cases, the redshift distribution rises with
redshift as more volume becomes available, but it drops sharply
at $r_0$ for the volume-limited survey, 
while the tail of the flux-limited survey extends to higher redshift $z\geq1$.
By defining the spherical Fourier modes through Eq.~(\ref{eq:klmkey}), 
we are implicitly assuming that the galaxy overdensity $\delta_g(\xvec)$ 
is pixelized. Therefore, the survey window function $\RW(r)$ should in 
principle include a pixel window function. For simplicity, however, we
will ignore this complication and assume that the data has been 
deconvolved.

Figure~\ref{fig:gaussian} illustrates the spherical power spectra of 
representative components of the full general relativistic formula
$\delta_g^\up{obs}$ in Eq.~(\ref{eq:fullGR}),
such as the matter density~$\dm$, 
the line-of-sight velocity~$V$, the gravitational potential~$\phi$, and so on
(see  Eqs.~[\ref{eq:los}]$-$[\ref{eq:zdist}]). The Gaussian window function
is adopted in computing the spherical power spectra.
For simplicity, we first assume that the matter transfer function $\TTm(k,z)$
is independent of redshift (i.e., of the line-of-sight 
distance~$\tilde r$) and the redshift-dependent quantities in the weight 
functions $\WW_l^i(r,\tilde r,k)$
such as $a$, $H$, and~$f$ are set equal to those evaluated at $z=0$
when computing the spherical multipole functions $\ST_l^i(\tilde k,k)$ in
Eq.~(\ref{eq:MM}). The spherical multipole function is then integrated 
over~$\tilde k$ to obtain the spherical power spectrum $\SP_l(k)$ by using
Eq.~(\ref{eq:sffp}). The full spherical galaxy power spectrum 
$\SP_l^\up{obs}(k)$ is the sum of the auto and the cross power spectra of 
all the components in Eq.~(\ref{eq:fullGR}), which can be readily deduced
from Fig.~\ref{fig:gaussian} (see also \cite{YOO10}).

The first column shows the spherical power spectra $\SP_l(k)$ 
of the matter density~$\dm$
(solid) and the redshift-space distortion 
${1+z\over H}{\partial V\over\partial r}$
(dashed) as a function of wavenumber~$k$ and angular multipole~$l$.
Each row shows the same spherical
power spectra but at different angular multipoles, ranging from $l=2$ (top)
to $l=10$ (bottom). Since the matter density power spectrum $\PM(k)$
is isotropic, the spherical power spectrum $\SP_l^\delta(k)$ (solid) of
the matter density is independent of angular multipole,
and it reduces to the three-dimensional power spectrum 
$\PM(k)\simeq\SP_l^\delta(k)$ (gray solid).
Given angular multipole~$l$ in each
row, the characteristic wavenumber $k_c\equiv l/r_0$ is shown as the vertical
lines, below which the power is highly suppressed and well beyond which the
equality of two power spectra holds. Technically, the power is suppressed
at $k\ll k_c$ due to the spherical Bessel function $j_l(kr)\simeq0$ 
at $kr\ll l$, and the physical interpretation is that larger scale modes 
$k\ll k_c\simeq k_\perp$ cannot be measured given the survey depth $r_0$.
The match between $\SP_l^{\delta}(k)$ and $\PM(k)$ could be extended to
larger scales, if more survey volume is available than what is assumed here.

For the Gaussian survey window function we adopted here, there exists an
analytic formula for the spherical multipole function $\ST_l^\delta(k',k)$.
Using Eq.~(6.633) in \citet{GRRY07} and setting $\lambda=0$ and $\mu=\nu=l$,
a useful identity can be derived as
\beeq
\label{eq:modi}
\int_0^\infty dx~x^2 e^{-\alpha x^2}j_l(\beta x)
j_l(\gamma x)={\pi e^{-{\beta^2+\gamma^2\over4\alpha}}
\over4\alpha\sqrt{\beta\gamma}}
I_{l+{1\over2}}\left({\beta\gamma\over2\alpha}\right)~,
\eneq
where the modified Bessel function is $I_\nu(z)=(-i)^\nu J_\nu(iz)$.
If we again ignore the time evolution of the transfer function
$\TTm(k,z)\simeq\TTm(k)$, the spherical multipole function of the matter
density is \cite{RARE12}
\beeq
\hspace{-5pt}
\ST^\delta_l(k',k)=\sqrt{\pi\over2k^{\prime2}}\TTm(k')
{r_0^2\sqrt{kk'}\over2}~e^{-{(k^2+k^{\prime2})r_0^2\over4}}
~I_{l+{1\over2}}\left({r_0^2kk'\over2}\right)~,
\eneq
and in the limit $r_0^2kk'\gg2l(l+1)$ (valid at small scales)
it reduces to
\beeq
\ST^\delta_l(k',k)=\sqrt{\pi\over2k^{\prime2}}~\TTm(k')\times
\left[{r_0\over\sqrt{4\pi}}~e^{-{(k-k')^2\over4}r_0^2}\right]~,
\label{eq:mulim}
\eneq
where the Gaussian wavepacket in the square bracket
converges towards the Dirac distribution in the limit of 
infinite volume ($r_0\RA\infty$). The 
resulting spherical multipole function is independent of angular multipole~$l$,
reflecting the fact that high-$k$ modes do not feel the $r$-dependence of the 
Gaussian window function.

At sufficiently high~$k$, where $\ST^\delta_l(k',k)$ is well described by
Eq.~(\ref{eq:mulim}), the spherical power spectrum in Eq.~(\ref{eq:sffp}) 
becomes
\bear
\label{eq:sss}
\SP_l^\delta(k,k)&\simeq&\int d\tilde k~\PM(\tilde k)\left[
{r_0\over\sqrt{4\pi}}~e^{-{(k-\tilde k)^2\over4}r_0^2}
\right]^2~ \nonumber \\
&\simeq&\PM(k)~{r_0\over2\sqrt{2\pi}}~,
\enar
and we define the three-dimensional spherical power spectrum as
\beeq
\SP^\delta_l(k)=\SP^\delta_l(k,k)\left({r_0\over2\sqrt{2\pi}}\right)^{-1}~.
\eneq
Equation~(\ref{eq:sss}) reduces to
Eq.~(\ref{eq:ddd}), if the volume is infinite.
Since numerical integrations of the
spherical power spectrum in Eq.~(\ref{eq:sffp}) require fine sampling and
prone to numerical errors due to the highly oscillating nature of the
spherical Bessel functions in~$\ST_l(\tilde k, k)$,
we tested our numerical calculations against
the analytic solution, and both results agree to high precision.

Next, the dashed curves in the first column represent the redshift-space
distortion contribution ${1+z\over H}{\partial V\over\partial r}$.
Since there is only one line-of-sight direction in the plane parallel limit,
the redshift-distortion contribution
is conveniently expressed in terms of cosine angle $\mu_k=\Xang\cdot\Kang$ 
between the line-of-sight $\Xang$ and the wavevector~$\kvec$,
and its three-dimensional power spectrum is $\PZ(k,\mu_k)= f^2\mu_k^4\PM(k)$.
Thin dashed curves show the spherical power spectrum $\SP_l^z(k)$
of the redshift-space distortion, while the gray dashed curves in each row
show $\PZ(k,\mu_k=1)=f^2\PM(k)$ for comparison. As the angular multipole
becomes higher at a given~$k=|\kvec|$, more angular modes are emphasized 
(or equivalently lower $\mu_k$), and the spherical power spectrum $\SP_l^z(k)$ 
decreases. At $k\gg k_c$, the spherical
power spectrum reduces to the three-dimensional power spectrum $\PZ(k,\mu_k)$
with its angular dependence encoded in angular multipole in each row.

The second column plots the spherical power spectra of the dominant
relativistic corrections to the observed galaxy fluctuation 
$\delta_g^\up{obs}$, such as the line-of-sight velocity~$V$ (solid),
the gravitational potential~$\phi$ (dashed), and their linear combinations
in Eqs.~(\ref{eq:los})$-$(\ref{eq:zdist}). Since the velocity and the 
potential contributions scale as $(k/\HH)$ and $(k/\HH)^2$ relative to 
the matter density~$\dm$, these relativistic contributions are comparable only
at the horizon scale $k\simeq\HH$ and constitute small corrections in most
galaxy surveys (see \cite{YOHAET12} for measuring these effects).

\begin{figure*}
\centerline{\psfig{file=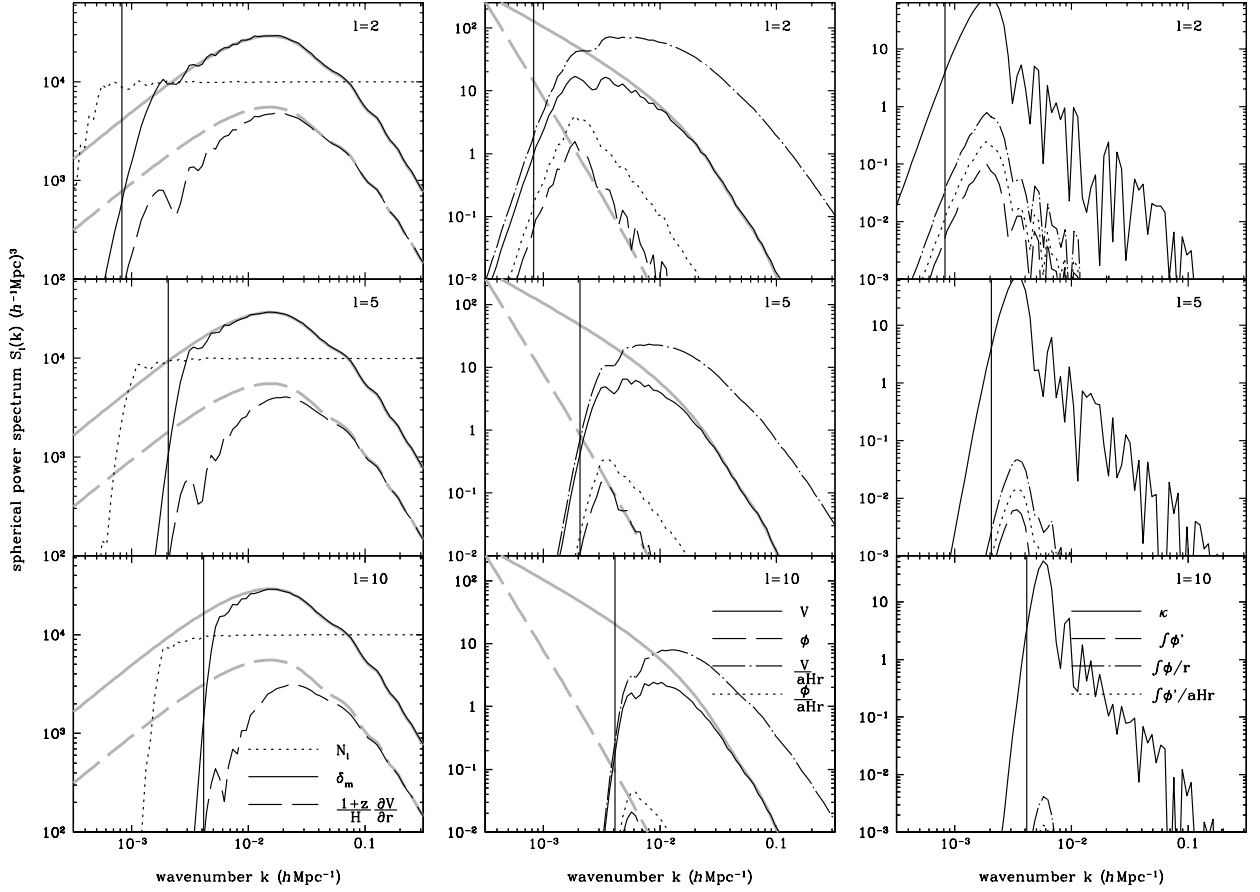, width=5.0in, angle=-90}}
\caption{Spherical power spectra with top-hat window function, in the
same format as in Fig.~\ref{fig:gaussian}. The general trend in the spherical
power spectra are similar, but oscillations are imprinted in the spherical
power spectra, due to the incomplete cancellation of the spherical Bessel
function, imposed by the top-hat window function.
The gray curves are identical to those in Fig.~\ref{fig:gaussian}. }
\label{fig:tophat}
\end{figure*}

The spherical power spectrum $\SP_l^\phi(k)$ (dashed) of the 
gravitational potential~$\phi$ follows the three 
dimensional power spectrum $P_\phi(k)$ (gray dashed) at each row, whose 
amplitude quickly decreases at high~$k$ as $P_\phi(k)\propto k^{-3}$, but
is independent of angular multipole~$l$.
At a given small~$k$, the amplitude $\SP_l^\phi(k)$ becomes lower in the 
lower panels, simply because $k\leq k_c$ 
(the vertical line in each row again illustrates
the characteristic minimum wavenumber~$k_c$ given angular multipole~$l$).
The angular dependence of the line-of-sight velocity~$V$ means that
its spherical power spectrum $\SP_l^V(k)$ (solid) also depends on angular
multipole~$l$ and it decreases with increasing angular multipole (or
smaller $\mu_k$) on all scales, in addition to the 
effect at $k\leq k_c$~.
Since the three dimensional power spectrum of the 
line-of-sight velocity is 
$P_V(k,\mu_k)=\mu_k^2P_v(k)=(\HH f/k)^2\mu_k^2\PM(k)$,
we plot $P_V(k,\mu_k=1)$ (gray solid) at each row for comparison, 
and again the spherical power spectrum $\SP_l^V(k)$ reduces to the
three-dimensional power spectrum $P_V(k,\mu_k)$.
Last, the dotted curves show the noise power spectrum $\NP_l(k)$ for a
galaxy sample with $\tilde n_g=10^{-4}~(\hmpc)^{-3}$.
For the Gaussian radial selection function, the shot-noise contribution to 
$\SP_l(k,k')$ can be computed exactly by using Eq.~(\ref{eq:modi}) as
\beeq
{\cal N}_l(k,k') =\frac{1}{2}{r_0^2\over\tilde n_g}
\sqrt{k k'} e^{-\frac{r_0^2}{4}(k^2+k'^2)} 
I_{l+\frac{1}{2}}\!\left(\frac{r_0^2 k k'}{2}\right)~,
\eneq
and it asymptotically matches the Poisson power spectrum $1/\tilde n_g$
on small scales.

The remaining two curves show the variants of the line-of-sight velocity 
and the gravitational potential, $V/\HH r$ (dot dashed) and $\phi/\HH r$ 
(dotted). They both arise due to the radial distortion $\drg/r$. 
With the prefactor $1/\HH r$ in those terms, more weight is given to shorter
distance~$r$, and hence they peak at a slightly higher wavenumber~$k$ than 
their counterparts without the prefactor, while the shift in the peak position
depends on the adopted window function.

The last column shows the spherical power spectra of the projected
quantities along the line-of-sight direction, such as the gravitational 
lensing~$\kag$ (solid), the integrated Sachs-Wolfe effect 
$\int_0^r d\tilde r~\phi'$ (dashed),
the Sachs-Wolfe effect $\int_0^r d\tilde r~\phi/r$ (dot-dashed), 
and their linear
combinations in Eqs.~(\ref{eq:los})$-$(\ref{eq:zdist}). Since 
these contributions are intrinsically angular and their variations are
largely limited to the transverse direction, the angular power spectrum
$C_l$ is often used to characterize them, and there is no exact analogy to
the usual three-dimensional power spectrum for the projected quantities
(see, however, \cite{HUGALO08}).
Apart from the gravitational lensing convergence, the spherical power spectra
of the projected quantities are at least an order-of-magnitude smaller
than the gravitational potential~$\phi$ in the second column.
Those projected quantities are isotropic and independent of angular multipole,
with the sole exception of the gravitational lensing~$\kag$, whose additional
angular dependence $l(l+1)$ arises due to the angular Laplacian operator.
The spherical power spectrum of the gravitational lensing $\kag$ exactly
corresponds to the 3D weak lensing \cite{HEAVE03,CAHEKI05,KIHEMI11}, 
where the same
spherical Fourier analysis is applied to background source galaxies to
map the foreground matter distribution. While two more shear fields are 
available in weak lensing measurements, only the convergence field~$\kag$
contributes to galaxy clustering.

\begin{figure*}
\centerline{\psfig{file=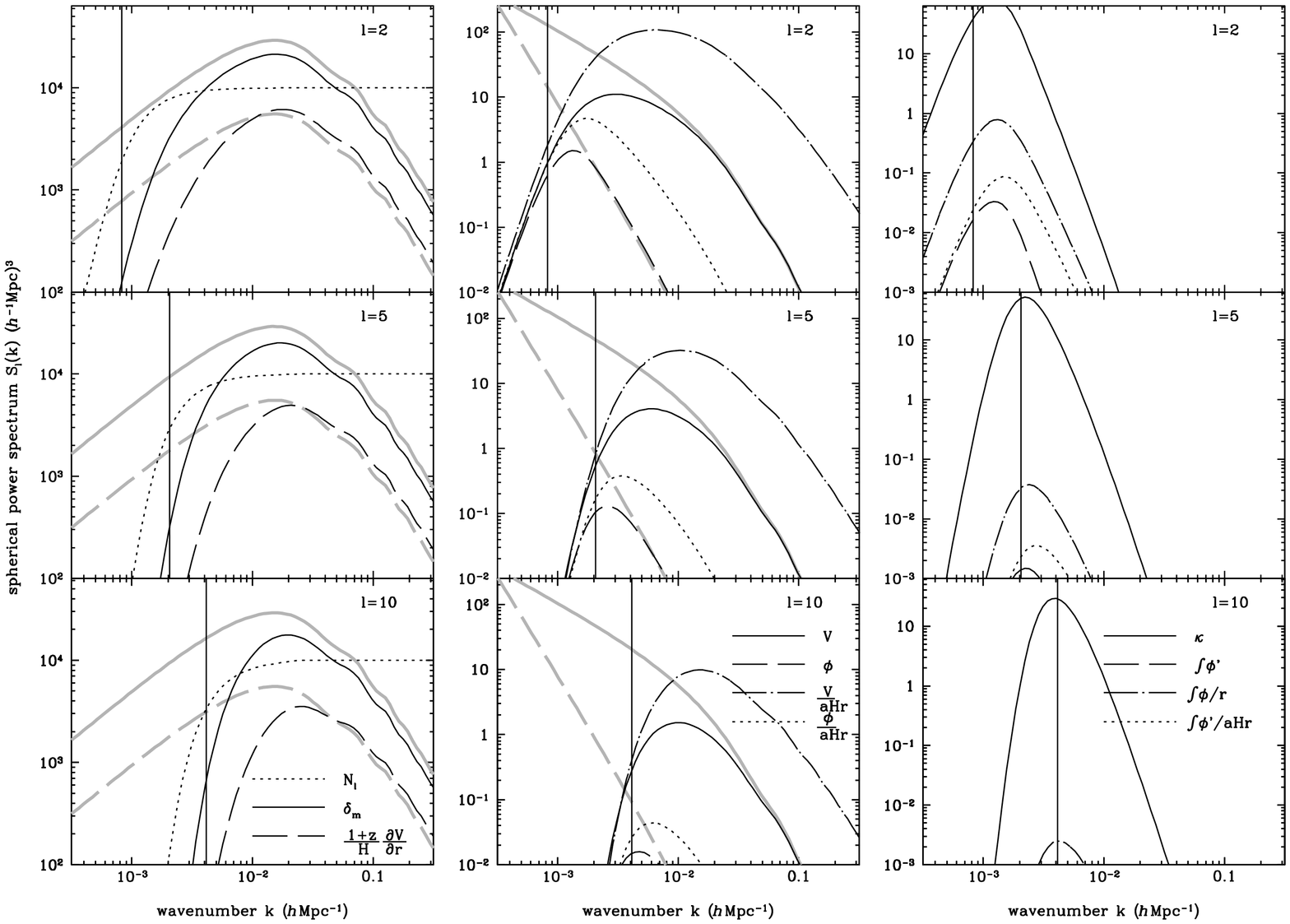, width=5.0in, angle=-90}}
\caption{Spherical power spectra with Gaussian window function. The 
time evolution
of the transfer and the weight functions over the line-of-sight distance
is explicitly computed. Compared to the spherical power spectra without
evolution in Fig.~\ref{fig:gaussian}, the spherical power spectra remain
largely unaffected in shape, while the amplitude is naturally lower than
the three-dimensional power spectra (gray) at $z=0$.
Various curves are in the same format as in Fig.~\ref{fig:gaussian}.}
\label{fig:evol}
\end{figure*}

In Fig.~\ref{fig:tophat}, we show the spherical power spectra with
the top-hat window function in Eq.~(\ref{eq:tophat})
and discuss its effect on the spherical power spectra, compared to those in
Fig.~\ref{fig:gaussian}. We have again ignored the time evolution of the 
transfer and the weight functions for simplicity. Three-dimensional power
spectra shown as gray curves in Fig.~\ref{fig:tophat} are identical to 
those in Fig.~\ref{fig:gaussian}.
The normalization of the spherical power spectra can be obtained in a similar
manner by considering the spherical power spectrum of the matter density,
\beeq
\SP_l^\delta(k,k)=\int d\tilde k~\PM(\tilde k)\left[{2~k\tilde k\over\pi}
\int_0^{r_0}dr~r^2~j_l(kr)j_l(\tilde kr)\right]^2~,
\eneq
and taking the limit $k\RA\infty$ (see Appendix~\ref{app:nor})
\beeq
\lim_{k\RA\infty}\SP_l^\delta(k,k)=\PM(k){r_0\over\pi}\sqrt{3\over\pi}~.
\label{eq:topnor}
\eneq
Therefore, we plot the spherical power spectra in Fig.~\ref{fig:tophat} as
\beeq
\SP_l(k)=\SP_l(k,k)\left({r_0\over\pi}\sqrt{3\over\pi}\right)^{-1}~.
\eneq

The spherical power spectra of the matter density (solid) and
the redshift-space distortion (dashed) in the first column are nearly identical
to their counterparts with the Gaussian window function. The prominent
difference is the oscillations in $\SP_l(k)$ at $k\leq2\pi/(r_0/3)$, where
most contributions to $\SP_l(k)$ at those low~$k$ arise near $r\simeq r_0$
and the highly oscillating spherical Bessel function is abruptly truncated
at the survey boundary $r=r_0$, causing incomplete cancellation of the
oscillations. The noise power spectrum (dotted) is nearly constant under the
top-hat survey window function.
With full weight given $\RW(r)=1$ near $r\leq r_0$, however,
the recovery of the three-dimensional power spectra is somewhat extended to
lower~$k$ than in Fig.~\ref{fig:gaussian}.

A similar pattern is observed for the line-of-sight velocity $\SP_l^V(k)$
and the potential $\SP_l^\phi(k)$ spherical power spectra in the second column.
Due to additional suppression factor $1/k^2$ in the weight function for
the potential~$\phi$, more weight is given to large~$r$ at a given~$k$ 
for~$\phi$ than for~$V$, which results in oscillations in $\SP_l^\phi(k)$
even at $k>2\pi/(r_0/3)$. The remaining two curves (dot-dashed and dotted) in 
the radial distortion follow the same trend but, compared to 
Fig.~\ref{fig:gaussian},
their peak is somewhat shifted to larger scale~$k$ due to the change in the 
window function.

The spherical power spectra of the projected quantities in the third column
show the reversed trend,  smooth power on large scales and oscillations on
small scales, compared to the spherical power spectra in the first column.
The projected quantities have additional integral in the spherical multipole
function $\ST_l(\tilde k,k)$, 
and this line-of-sight integration cancels the oscillating part
of the second spherical Bessel function, while the overall oscillation is
determined by the first spherical Bessel function. The high-$k$ oscillation 
arises near $r\gg l/k$, at which there is no suppression in the top-hat window
function. Furthermore, the peak position is shifted to smaller scales,
since the mean distance to the source galaxies is smaller with the top-hat
window function.

Having understood the key features in the spherical power spectra
with the simplifying assumptions, we are now in a position to consider the
full generality: time dependence of the transfer function and the weight
function on the line-of-sight distance.
In Fig.~\ref{fig:evol}, we present the spherical power spectra
with the Gaussian window function, explicitly accounting for their time
dependence in the spherical multipole function. Overall, the spherical power 
spectra in Fig.~\ref{fig:evol} resemble those in Fig.~\ref{fig:gaussian},
where the time dependence is neglected. The key difference in this case
lies in the amplitude of the spherical power spectra, as the matter transfer
function $\TTm(k,r)$ decreases in amplitude at $r>0$ (i.e., $z>0$).

The first column shows the spherical power spectra of the matter density
(solid) and the redshift-space distortion (dashed). Compared to the flat-sky
matter power spectrum (gray solid) at $z=0$, the spherical power spectrum
$\SP_l^\delta(k)$ is lower in amplitude, as it is obtained by averaging
over a range of redshift in Fig.~\ref{fig:window}. For the same reason,
the logarithmic growth rate~$f$ is higher at $z>0$, and hence the spherical
power spectrum of the redshift-space distortion is higher in amplitude.
The noise power spectrum (dotted) remains unchanged in Fig.~\ref{fig:evol},
since the time dependence of the mean number density is already
accounted by the survey window function in Fig.~\ref{fig:gaussian}.

In the second column, the spherical power spectra of the line-of-sight 
velocity, the gravitational potential, and their variants show little 
difference, 
compared to those in Fig.~\ref{fig:gaussian}. A small change in amplitude
arises due to the shift in the peak line-of-sight distance. 
The velocity is nearly constant
in redshift, while the gravitational potential is somewhat larger at $z>0$,
as dark energy domination leads to the decay in the gravitational potential.
This trend is further reflected in the third column, where the projected 
quantities are displayed.
The spherical power spectra of the gravitational
lensing (solid) and the gravitational potential (dot-dashed) are slightly
larger in amplitude, while the integrated Sachs-Wolfe contributions 
(dashed and dotted) are smaller, since the potential is constant in the
Einstein-de~Sitter phase at $z>0$.

\subsection{Statistical uncertainties}
\label{ssec:uncer}
The covariance matrix of the spherical power spectrum $\SP_l(k,k)$ generally 
involves the power spectrum, bispectrum and trispectrum of the spherical 
Fourier modes $\delta_{lm}(k)$. Non-Gaussianity induced by nonlinear 
gravitational clustering or, possibly, already present in the initial 
conditions can generate significant covariance through the bispectrum and 
trispectrum. 
For simplicity, however, we will henceforth ignore these contributions and 
assume that $\delta_{lm}(k)$ is simply Gaussian.

We begin with the calculation of the covariance of the spherical power 
spectrum assuming an all-sky experiment. At fixed multipole $l$, $\SP_l(k,k)$ 
can be estimated by 
averaging over all $m$ and a thin wavenumber interval $\Delta k$ and by
subtracting the shot-noise component.
However, since the shot-noise contribution 
$\NP_l$ is difficult to compute without precise knowledge of the radial 
selection function, we simply present the observed spherical power spectrum,
i.e., without shot-noise subtraction. We thus define 
an estimator of the spherical power spectrum
\beeq
\hat{\SP}_l(k,k)=\frac{1}{2 l+1} \sum_m \frac{1}{\Delta k}
\int_{_{\Delta k}}\!\!du\,|\delta_{l m}(u)|^2 ~,
\eneq
where the integration runs over the domain $[k-\Delta k/2,k+\Delta k/2$,
given a chosen band width $\Delta k$. The 
covariance of this (unbiased) band-power estimator is
\begin{widetext}
\beeq
{\rm Cov}[\hat{\SP}_l(k,k),\hat{\SP}_{l'}(k',k')] =
\frac{1}{\left(2 l+1\right)\left(2 l'+1\right)}
\sum_{m,m'}\frac{1}{(\Delta k)^2}\int_{_{\Delta k}}\!\!du\int_{_{\Delta k}}\!\!du'\,
\Bigl[\left\langle|\delta_{l m}(u)|^2|\delta_{l' m'}(u')|^2\right\rangle
-\left\langle|\delta_{l m}(u)|^2\right\rangle 
\left\langle|\delta_{l'm'}(u')|^2\right\rangle\Bigr] ~.
\label{eq:cov1}
\eneq
The assumption of Gaussian random fields considerably simplifies the evaluation
of the trispectrum. A straightforward application of Wick's theorem yields
\bear
{\rm Cov}[\hat{\SP}_l(k,k),\hat{\SP}_{l'}(k',k')] &=&
\frac{1}{\left(2 l+1\right)^2}
\sum_{m,m'}\frac{1}{(\Delta k)^2} \int_{_{\Delta k}}\!\!du 
\int_{_{\Delta k}}\!\!du'\, \Bigl[\bar{\SP}_l(u,u')+\NP_l(u,u')\Bigr]^2 
\bigl(\delta_{m, m'}+\delta_{m,-m'}\bigr) \delta_{l l'} \nonumber \\
&=&\frac{2}{2 l+1}\frac{1}{(\Delta k)^2}
\int_{_{\Delta k}}\!\!du\int_{_{\Delta k}}\!\!du'\, 
\Bigl[\bar{\SP}_l(u,u')+\NP_l(u,u')\Bigr]^2 
\delta_{ll'} ~.
\label{eq:cov2}
\enar
\end{widetext}
In the flat-sky approximation, estimates from different band-powers are 
uncorrelated in the Gaussian limit, so that the covariance of power spectrum 
estimators Cov$[\hat P(k_i),\hat P(k_j)]$ is proportional to 
$2 P(k_i) P(k_j) \delta_{k_i,k_j}$.
In the spherical Fourier decomposition, the Limber approximation shows that, 
for the matter (and potential) perturbations, band-power estimates of the 
spherical power spectrum are also uncorrelated. However, contributions from,
e.g., the line-of-sight velocity or z-distortions will introduce covariance 
between measurements from different band-powers.

For shorthand convenience, let us now define
\beeq
\SP_l(k)\equiv \bar{\SP}_l(k) + \NP_l(k)~,
\eneq
If we restrict ourselves to a galaxy sample of constant comoving density 
$\bar n_g=\tilde n_g = N_s/V_s$ and Poisson white noise, the error on 
$\hat{\SP}_l(k)$ scales as
\beeq
\bigl[\Delta\hat{\SP}_l(k)\bigr]^2 = \frac{2}{2 l+1} 
\left[\bar{\SP}_l(k)+\frac{1}{\bar{n}_g}\right]^2 
\label{eq:lcov}
\eneq
for infinitely narrow $k$ bins ($\Delta k\rightarrow0$).
Equation~(\ref{eq:lcov}) is the intrinsic error in $\hat\SP_l(k)$,
reduced by the number of angular mode~$l$ available in the all-sky survey 
volume. This expression is a 2D analogy
to the FKP expression \cite{FEKAPE94} in 3D, 
except that the
number $N(k)/2$ of independent modes in a $k$-shell is replaced by the number
$(2l+1)/2$ of independent $\ket{klm}$ modes at fixed $k$ and $l$. 
In general, however, the survey window function is not uniform and $1/\bar n_g$
must be replaced by the line-of-sight integral $\NP_l(k)$.
In addition to angular modes, there exist more radial modes available in the
survey.
The cumulative signal-to-noise for a measurement of $\bar{\SP}_l$ is then
\beeq
\hspace{-10pt}
\left(\frac{S}{N}\right)^2 = \frac{V_s^{1/3}}{2}
\sum_{l=2}^{\lmax} \left({2l+1\over2}\right)\int_{\kmin}^{\kmax}
\!\!\frac{dk}{2\pi}\,
\left(\frac{\bar{\SP}_l(k)}{\NP_l(k)+\bar{\SP}_l(k)}\right)^2~,
\eneq
where $V_s^{1/3}k$ is the number of radial modes at~$k$.
We introduced an extra multiplicative factor of $V_s^{1/3}$ (which is the 
only scale in the problem) in order to get a dimensionless quantity. This step
can be made more rigorous by considering pixelized window functions along the 
line of sight. Going into this level of details is, however, beyond the scope
of this work.
The factor of $1/2$ arises from 
the fact that $\delta(\xvec)$ is real, i.e. $\delta_{lm}^*(k)
=\delta_{l -m}(k)$.
Note that the cumulative signal-to-noise for a measurement of, e.g., the 
spherical power induced by the line-of-sight velocity is given by the same 
formula with $\SP_l^V$ replacing $\bar{\SP}_l$ in the numerator.

\begin{figure}
\centerline{\psfig{file=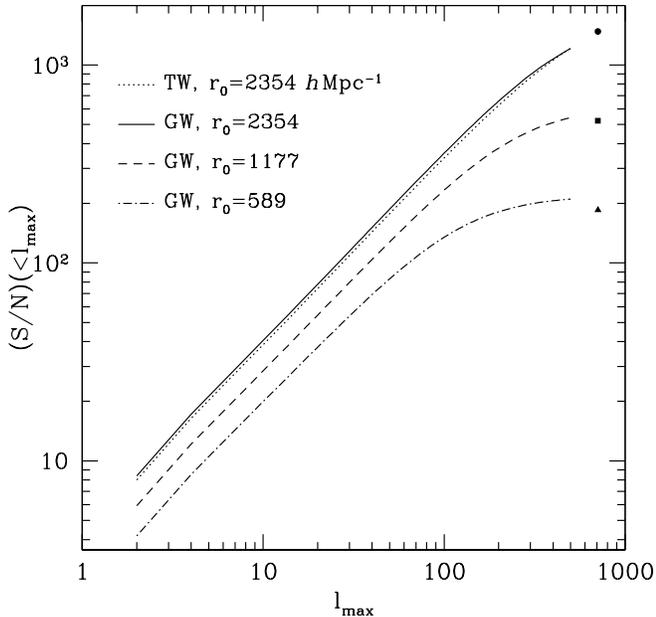, width=3.4in}}
\caption{Signal-to-noise for a measurement of the spherical
power spectrum of galaxies with constant bias $b=2$ (see text). Results are 
shown as a function of the maximum multipole $l_{\rm max}$. The curves show 
Eq.~(\ref{eq:limbersn}) for the tophat (TW) and Gaussian (GW) survey window
with several choices of $r_0$. The filled symbols show $S/N$ in the flat-sky
approximation (Eq.~[\ref{eq:flatskysn}]) for the Gaussian survey window with 
$r_0=2354$ (circle), 1177 (square) and 589~$\hmpc$ (triangle). Note that their 
abscisse is arbitrary.}
\label{fig:sn}
\end{figure}

In order to establish a correspondence with mode counting in the 3D flat case, 
let us consider the limit $\bar{\SP}_l\gg \NP_l$. 
Let also $L = V_s^{1/3}$ and $\Delta L$ be the characteristic length and 
resolution of the survey. 
The maximum wavenumber and wavenumber resolution thus are 
$\kmax=2\pi/\Delta L$ and $\Delta k = 2\pi/ L$. 
The angular resolution of the survey is $\theta\sim \Delta L/L$, which 
yields a maximum multipole $\lmax\sim 1/\theta \sim \kmax/\Delta k$. 
Therefore, the signal-to-noise is
\bear
\left(\frac{S}{N}\right)^2 &\simeq& 
\frac{1}{4}~\lmax~ (\lmax+2)~\left(\frac{\kmax}{\Delta k}\right) \\
&\simeq& \frac{1}{4} \left(\frac{\kmax}{\Delta k}\right)^3 ~. \nonumber 
\enar
This is equal (up to a factor of order unity) to the signal-to-noise in the
flat-sky limit, 
\bear
\left({S\over N}\right)^2&=&{V_s\over2}\times2\pi\int 
{dk_\perp~k_\perp\over(2\pi)^2}\int {dk_\parallel\over2\pi} \\
&\simeq&{\pi\over2}\left({\kmax^2-\kmin^2\over\Delta k^2}\right) 
\left({\kmax-\kmin\over\Delta k}\right) \nonumber \\
&\simeq& \frac{\pi}{2}\left(\frac{\kmax}{\Delta k}\right)^3 ~, \nonumber
\enar
where the last equality assumes $\kmin\ll \kmax$. When dealing with real 
galaxy data however, one must take into account the shot-noise and the survey 
window function. Ignoring contributions others than fluctuations in the matter 
density, the Limber approximation gives
\bear
\left(\frac{S}{N}\right)^2_{l<\lmax} &\approx& \frac{V_s^{1/3}}{2}
\sum_{l=2}^{\lmax} \left({2l+1\over2}\right) \\ 
&& \times \int_0^\infty\!\!\frac{dk}{2\pi}~
\left(\frac{\tilde n_g\RW(\frac{\nu}{k}) P_m(k,\frac{\nu}{k})}
{1+\tilde n_g\RW(\frac{\nu}{k})P_m(k,\frac{\nu}{k})}\right)^2~.
\label{eq:limbersn}
\nonumber
\enar
In Fig. \ref{fig:sn}, this expression is evaluated as a function of $\lmax$ for
the Gaussian (GW) and tophat (TW) survey window considered in Sec.~\S\ref{sec:results} 
(dotted and solid curves). The galaxy bias is assumed to be constant, $b=2$, 
and the $z$-dependence of the matter power spectrum has been neglected, i.e., 
$P_m(k,r)=P_m(k,0)$. 
To exemplify the scaling $(S/N)\propto V_s^{1/6}$, results are also shown for the 
Gaussian window with 1/8 and 1/64 the fiducial volume $V_s\approx 55\hgpc$. 
The signal-to-noise eventually saturates at large values of $\lmax$ because, for 
$l\gg 1$, the Gaussian window $\RW(\nu/k)\sim \exp[-(l/k r_0)^2]$ only picks up
(small-scale) measurements that are shot-noise dominated. For our choice of $b$,
this occurs at $k\sim 0.2\hmpc$. As a result, the signal-to-noise flattens out  
around $\lmax\sim k r_0\sim$400 ($r_0=2354\hmpc$) and $\sim$100 ($r_0=589\hmpc$).
The filled symbols represent the signal-to-noise in the flat-sky approximation,
\beeq
\left(\frac{S}{N}\right)^2 =
\frac{1}{\pi}\int_0^\infty\!\!dr\,  r^2
\int_0^\infty\!\!dk\,  k^2 
\left(\frac{\tilde n_g\RW(r) P_m(k,r)}{1+\tilde n_g\RW(r)P_m(k,r)}\right)^2~,
\label{eq:flatskysn}
\eneq
computed for the Gaussian survey window with $r_0=2354$ (circle), 1177 (square)
and $589\hmpc$ (triangle). The flat-sky estimates are  consistent with the 
Limber-approximated $(S/N)(<\lmax)$ in the limit $\lmax\gg 1$. This provides 
support for the validity of Eq.~(\ref{eq:limbersn}).

\section{DISCUSSION}
\label{sec:discussion}
We have performed an all-sky analysis of the galaxy power spectrum, accounting
for all the relativistic effects in galaxy clustering. The spherical
Fourier analysis has been well developed in galaxy clustering
\cite{BIQU91,FILAET95,HETA95}, while its application to galaxy clustering
has been limited to the Kaiser formula \cite{KAISE87}. We have used the
spherical Fourier analysis to analyze the full relativistic formula.
The observed
galaxy fluctuation is decomposed in terms of spherical harmonics and
spherical Bessel functions that are angular and radial
eigenfunctions of the Helmholtz
equation, providing a natural basis for the observer at origin
to describe the galaxy clustering measurements on the observed sphere.

In light of the recent development in 
the relativistic formulation of galaxy clustering \cite{YOFIZA09,YOO10},
there exist numerous relativistic effects in galaxy clustering, in addition
to the standard redshift-space distortion effect. These relativistic effects
become substantial on very large scales, and measurements of these
large-scale modes inevitably invoke complications associated with
the flat-sky approximation and the survey geometries. 
By using the spherical harmonics for its angular decomposition,
the spherical power spectrum is independent of the validity of the flat-sky
approximation, while it retains the advantage of the standard Fourier 
analysis, namely, the simple and physically intuitive interpretation 
of the measurements in conjunction with the underlying matter distribution.

We have computed the spherical Fourier power spectrum of the observed
galaxy distribution, 
accounting for the relativistic effects. Compared to
the standard Newtonian description, there exist additional contributions to
the observed galaxy fluctuation, and these additional contributions can be 
categorized
as the matter density fluctuation, the line-of-sight velocity contribution,
the gravitational potential contribution, and the line-of-sight projection
contribution \cite{CHLE11,BODU11,BASEET11,BRCRET12,JESCHI12}.
The spherical power spectrum of the matter density is identical to the usual
three-dimensional matter power spectrum $\SP_l^\delta(k)\simeq P_m(k)$, 
regardless of its angular multipole~$l$, as the matter power spectrum is
isotropic. This correspondence greatly facilitates the physical interpretation
of the measurements.

Since the line-of-sight velocity affects the observed distance to the
galaxies in redshift space, the velocity contribution in galaxy clustering 
is angular dependent. Therefore, the spherical power spectrum $\SP_l^V(k)$
of the velocity contribution is similar to its three-dimensional counterpart
$P_V(k,\mu_k)$, but its angular dependence is encoded as a function of 
angular multipole~$l$:
At a given amplitude~$k$ of a wavevector~$\kvec$, higher angular multipoles
represent larger transverse modes, or lower cosine angle $\mu_k=k_\parallel/k$.
The redshift-space distortion effect arises from the spatial derivative of the
line-of-sight velocity, and its spherical power spectrum $\SP_l^z(k)$
follows the similar trend, as its three-dimesional
power spectrum is $P_z(k,\mu_k)=f^2\mu_k^4\PM(k)$.

The gravitational potential contribution to the observed galaxy fluctuation
can be readily computed from the spherical Fourier decomposition. 
Similar to the case of the matter density fluctuation, the gravitational
potential power spectrum $P_\phi(k)$ is isotropic, and 
the spherical power spectrum for the gravitational potential is identical
to the three-dimensional power spectrum $\SP_l^\phi(k)\simeq P_\phi(k)$,
regardless of its angular multipole. Furthermore, while the gravitational
potential contribution to the variance of galaxy clustering
often diverges due to its scale-free nature, the spherical power spectrum
is unaffected by this problem since it measures individual 
modes of fluctuations as in the traditional power spectrum analysis.

The other contributions to the observed galaxy fluctuation, 
such as the gravitational lensing and the integrated Sachs-Wolfe effects, 
arise from fluctuations along the line-of-sight direction.
Since these projected quantities are 
intrinsically angular, it is difficult to handle their contribution in the
standard power spectrum analysis.
However, with its angular decomposition using spherical harmonics, 
the spherical Fourier analysis can naturally
implement the contribution of the projected quantities to galaxy clustering.
With the sole exception of the gravitational lensing effect, we find that
compared to the matter density fluctuation, 
the contribution of the projected quantities are negligible, and 
this justifies the simplification of ignoring the projected quantities
in the power spectrum analysis \cite{YOO10,YOHAET12}.
Moreover, the spherical power spectrum $\SP_l^\kappa(k)$ of the gravitational
lensing contribution in galaxy clustering is also known as 3D weak lensing
\cite{HEAVE03,KIHEMI11}, where the information on radial distances to the
background source galaxies is utilized to map the matter distribution in 3D,
as opposed to the traditional 2D weak lensing. Our spherical Fourier
analysis provides a complete and comprehensive description of galaxy
clustering and its associated effects.

We have derived the covariance matrix of the spherical power spectrum,
assuming that the matter density fluctuation is the dominant contribution.
The covariance matrix of the spherical power spectrum
asymptotically matches that of the three-dimensional power spectrum
on small scales. It is also shown \cite{YOO10,YOHAET12,YOSE13} 
that since the volume available for galaxy surveys at low redshift
is relatively small, there are too few large-scale modes that are sensitive 
to the relativistic effect in galaxy clustering. Therefore, it makes little
difference in terms of measurement significance,
if one chooses to embed the observed sphere in a cubic volume
and performs the standard Fourier analysis, instead of performing 
the spherical Fourier
analysis. However, the spherical Fourier analysis presented in this paper
provides a more natural way to analyze the full relativistic effects in
galaxy clustering. Furthermore, it is shown \cite{YOHAET12} that
the multi-tracer
analysis \cite{SELJA09} with the shot-noise cancelling technique
\cite{SEHADE09} can substantially enhance the measurement significance
of the relativistic effects in galaxy clustering, in which case we expect
that the spherical Fourier analysis becomes essential in describing the
measurements on large scales.

\acknowledgments
We acknowledge useful discussions with Ruth Durrer, Ue-Li Pen, 
Uro{\v s} Seljak,  Zvonimir Vlah.
J.Y. is supported by the SNF Ambizione Grant.
V.D. acknowledges support from the Swiss National Science Foundation and is 
grateful for the hospitality of the Aspen Center for Physics, where part of 
this work was completed.

\bibliography{ms.bbl}

\appendix

\section{Top-hat normalization}
\label{app:nor}
Here we derive the normalization coefficient for the top-hat window function
in Eq.~(\ref{eq:topnor}). The normalization coefficient can be obtained
by computing the spherical power spectrum for the matter density
\beeq
\SP_l^\delta(k,k)=\int d\tilde k~\PM(\tilde k)\left[{2~k\tilde k\over\pi}
\int_0^{r_0}dr~r^2~j_l(kr)j_l(\tilde kr)\right]^2~,
\label{aeq:one}
\eneq
and taking the limit ($k\RA\infty$). We first define the integrand
\beeq
F(k,\tilde k)\equiv{2~k\tilde k\over\pi}
\int_0^{r_0}dr~r^2~j_l(kr)j_l(\tilde kr)~,
\eneq
and then arrange Eq.~(\ref{aeq:one}) as
\beeq
\SP_l^\delta(k,k)=\int_{-\infty}^\infty d\bar k~\PM(k+\bar k)\left[
F(k,k+\bar k)\right]^2~.
\eneq
Since the integrand peaks around $k\simeq\bar k$, we take the limit 
($k\RA\infty$) and expand the integrand $F(k,k+\bar k)$:
\bear
\lim_{k\RA\infty}\SP_l^\delta(k,k)&\simeq&
\PM(k)\int_{-\infty}^\infty d\bar k\left[
\FF+{\bar k^2\over2}~\FF''+\cdots\right]^2 \nonumber \\
&=&\PM(k)~\sqrt{2\pi}~\FF^2\left[-{\FF\over2\FF''}\right]^{1/2}~,
\enar
where we performed a Gaussian integral. The asymptotic values of the
integrand are
\bear
\FF&\equiv&\lim_{k\RA\infty}F(k)=\lim_{k\RA\infty}
{2r_0\over \pi x}\int_0^x dx~x^2~j_l^2(x)={r_0\over\pi}~,\\
\FF''&\equiv&\lim_{k\RA\infty}F''(k)=\lim_{k\RA\infty}
{2r_0^3\over \pi x^3}\int_0^x dx~x^4~j_l(x)\left[{2\over x}j_l'(x)+j_l''(x)
\right] \nonumber\\
&=&-{r_0^3\over3\pi}~,
\enar
and we obtain the normalization coefficient in Eq.~(\ref{eq:topnor})
\beeq
\lim_{k\RA\infty}\SP_l^\delta(k,k)=\PM(k){r_0\over\pi}\sqrt{3\over\pi}~.
\eneq

\section{Covariance of the spherical power spectrum}
\label{app:cov}
Here we present details of the calculation of the covariance matrix of  the
spherical power spectrum estimator. We will omit the band-power averaging for
the sake of conciseness. Using Eq.~(\ref{eq:klmkey}), 
the four-point correlator of the spherical Fourier modes reads
\begin{widetext}
\bear
\AVE{\delta_{lm}(k)\delta_{lm}^*(k)\delta_{l'm'}(k')\delta_{l'm'}^*(k')}
&=& 
\left(\frac{2}{\pi}\right)^2 k^2 k'^2
\prod_{i=1}^4\biggl\{\int\!\!d^3x_i\biggr\}
j_l(kr_1) j_l(k r_2) j_l(k' r_3) j_l(k' r_4) \nonumber \\
&& \times 
Y_{lm}^*(\Xang_1)Y_{lm}(\Xang_2) Y_{l'm'}^*(\Xang_3)Y_{l'm'}(\Xang_4)
\AVE{\delta(\xvec_1)\delta(\xvec_2)\delta(\xvec_3)\delta(\xvec_4)}~.
\enar
Assuming the galaxy overdensity field $\delta(\xvec)$ follows Gaussian statistics, 
the four-point correlator in the right-hand side reduces to the sum of three products 
of two-point correlation functions, whose explicit expression is given by 
Eq.~(\ref{eq:2ptngx}) divided by $\tilde n_g^2$. The contribution that involves 
$\AVE{\delta_g(\xvec_1)\delta_g(\xvec_2)}\AVE{\delta_g(\xvec_3)\delta_g(\xvec_4)}$
exactly cancels out the term $\AVE{|\delta_{lm}(k)|^2}\AVE{|\delta_{l'm'}(k')|^2}$
in the covariance matrix in Eq.~(\ref{eq:cov1}). 
To compute the two other contributions,
we must evaluate, e.g.,
\bear
J_{14} &\equiv& \frac{2}{\pi} k k'
\int\!\!dr_1 ~ r_1^2\int\!\!dr_4 ~r_4^2~ j_l(kr_1) j_{l'}(k' r_4)\int\!\!d^2\Xang_1
\int\!\!d^2\Xang_4~Y_{lm}^*(\Xang_1) Y_{l'm'}(\Xang_4)
\nonumber \\
&& \times 
\left\{\RW(r_1)\RW(r_4)\Bigl[1+\xi_g(\xvec_1-\xvec_4)\Bigr]+\frac{1}{\tilde n_g}
\RW(r_1)\delta^D(\xvec_1-\xvec_4)\right\} ~.
\enar
On expressing $\xi_g(\xvec_1-\xvec_4)$ as the Fourier transform
\beeq
\xi_g(\xvec_1-\xvec_4)= \int\!\!\frac{d^3k}{(2\pi)^3}
\TT_g(k,r_1) \TT_g(k,r_4) P_{\cp}(k) ~e^{i\kvec\cdot(\xvec_1-\xvec_4)} ~,
\eneq
and inserting the Rayleigh expansion in Eq.~(\ref{eq:rayleigh}), the integrals over the
angular variables simplify to
\beeq
\int\!\!d^2\Xang_1\int\!\!d^2\Xang_4~Y_{lm}^*(\Xang_1) Y_{l'm'}(\Xang_4)~
e^{i\kvec\cdot(\xvec_1-\xvec_4)}=
(4\pi)^2 i^{l-l'} j_l(k r_1) j_{l'}(k r_4) Y_{lm}^*(\Kang) Y_{l'm'}(\Kang) ~.
\eneq
As a consequence, the contribution of $\xi_g(\xvec_1-\xvec_4)$ to $J_{14}$ becomes
\bear
\lefteqn{32\pi k k' i^{l-l'}\int\!\!dr_1 r_1^2 \int\!\!dr_4 r_4^2 ~
\RW(r_1)\RW(r_4) j_l(k r_1) j_{l'}(k' r_4)
\int\!\!\frac{d^3\tilde k}{(2\pi)^3} 
\TT_g(\tilde k,r_1) \TT_g(\tilde k,r_4) j_l(\tilde k r_1) j_{l'}(\tilde k r_4)
Y_{lm}^*(\hat{\tilde\kvec}) Y_{l'm'}(\hat{\tilde\kvec}) P_{\cp}(\tilde k)} && 
\nonumber \\
&& = (4\pi)\int\!\!d\ln\tilde k ~\Delta_{\cp}^2\!(\tilde k)
\left[
\frac{2}{\pi}k k'
\int\!\!dr_1 r_1^2 \RW(r_1)j_l(k r_1) j_l(\tilde k r_1) \TT_g(\tilde k,r_1)
\times
\int\!\!dr_4 r_4^2 \RW(r_4)j_l(k r_4) j_l(\tilde k r_4) \TT_g(\tilde k,r_4)
\right] \delta_{l l'}\delta_{m m'}\nonumber \\
 && = \SP_l(k,k')~ \delta_{l l'} \delta_{m m'} ~.
\enar
Similarly, using 
$\delta^D(\xvec_1-\xvec_4)=\delta^D(r_1-r_4)\delta^D(\Xang_1-\Xang_4)/r_1^2$ 
and the 
orthonormality of the spherical harmonics, the contribution of the Poisson noise 
term to $J_{14}$ is
\beeq
\frac{2k k'}{\pi\tilde n_g} \int\!\!dr_1 r_1^2 \int\!\! dr_4 r_4^2
\RW(r_1) j_l(kr_1) j_{l'}(k' r_4) \delta^D(r_1-r_4) 
\int\!\!d^2\Xang_1 Y_{lm}^*(\Xang_1) Y_{l'm'}(\Xang_1)
= \NP_l(k,k') ~ \delta_{ll'} \delta_{mm'} ~.
\eneq
$J_{14}$ is the sum of these two contributions, i.e. 
$J_{14}=[\SP_l(k,k')+\NP_l(k,k')]\delta_{ll'}\delta_{mm'}$. 
Symmetry considerations 
show that this holds true for $J_{ij}$, $i\ne j$. This leads to 
the desired result in Eq.~(\ref{eq:cov2}).
\end{widetext}

\end{document}